\documentclass[final,5p,times,twocolumn]{elsarticle}

\usepackage{amssymb}
\usepackage{amsmath}
\usepackage{comment}
\usepackage{amsthm}
\usepackage{xurl}
\usepackage[colorlinks,allcolors=blue]{hyperref}
\usepackage[T1]{fontenc}
\usepackage[utf8]{inputenc} 
\usepackage{xcolor}

\usepackage{lineno}
\journal{Astroparticle Physics}

\begin{document}
\begin{frontmatter}



\title{Simulations of Astrophysically Relevant Pair Beam Instabilities in a Laboratory Context}


\author[label1]{Suman Dey\corref{cor1}}
\ead{suman.dey@desy.de}
\author[label1]{G\"{u}nter Sigl} 

\cortext[cor1]{Corresponding author}
\affiliation[label1]{organization={II. Institut f\"{u}r Theoretische Physik, Universit\"{a}t Hamburg},
            addressline={Luruper Chaussee 149}, 
            city={Hamburg},
            postcode={22761},
            country={Germany}}

\begin{abstract}
The interaction of TeV blazars emitted gamma-rays with the extragalactic background photons gives rise to a relativistic beam of electron-positron ($e^- e^+$) pairs propagating through the intergalactic medium, producing a cascade through up-scattering low-energy photons. Plasma instability is considered one of the underlying energy-loss processes of the beams. We employ particle-in-cell (PIC) simulations to study the plasma instabilities of relativistic pair beams propagating in a denser background plasma, using the parameters designed to replicate astrophysical jets under laboratory conditions. In an astrophysical scenario with a broad, dilute beam, electromagnetic instability is suppressed because the beam exhibits momentum anisotropy with a large longitudinal momentum spread compared to its transverse momentum. We find the range of density contrast at which electrostatic modes are dominating over electromagnetic modes with an anisotropic beam in laboratory scales, consistent with the physically relevant conditions for Blazar-induced beams.
We have used a broad Cauchy distribution for the beam particles, which is more realistic in representing the non-Maxwellian nature of pair beams, improving upon previous studies. We investigate the interplay between the instability-generated magnetic field and the momentum anisotropy of the beam. We extrapolate the beam energy loss and the angular broadening due to non-linear feedback of instability. We find that the astrophysical beams have lost approximately 4\% of their total energy due to instability. Nevertheless, the instability generates a negligible angular broadening for Blazar-induced beams.
\end{abstract}





\end{frontmatter}


\section{Introduction}\label{sec:intro}
Blazars are a type of active galactic nuclei (AGNs) featuring jets of high-energy (i.e. $E \geq 100$ MeV) particles that are oriented almost directly toward Earth. The primary gamma rays with TeV energies travel through the intergalactic medium and interact with the extragalactic background light (EBL). This interaction leads to attenuation of the primary TeV photons, especially for distant blazars. The relativistic $e^- e^+$ pair plasmas are produced when the TeV gamma-ray interacts with a low-energy EBL photon. The relativistic pairs then undergo inverse Compton (IC) scattering with cosmic microwave background (CMB) photons \cite{gould1967opacity, blumenthal1970bremsstrahlung}. This cycle of pair production and IC scattering continues, forming an extended electromagnetic cascade of secondary GeV gamma rays. Nevertheless, there is a disagreement between the expected \cite{aharonian2001tev} and observational photon spectra measured from Fermi-LAT and imaging atmospheric (or air) Cherenkov telescopes (e.g., MAGIC, VERITAS, and HESS) \cite{neronov2009sensitivity, neronov2010evidence}, known as GeV-TeV tension. One potential explanation for this missing GeV cascade emission can be understood as the deflection of the pairs by the intergalactic magnetic fields (IGMF). This deflection leads to a time delay of the cascade photons. In addition, the deflected particles create extended GeV emission around the blazar, which can extend beyond the field of view of the detector. Consequently, the non-observation of extended GeV emission spectra can be used to estimate lower bounds on the IGMF
strength \cite{Elyiv:2009bx, neronov2009sensitivity, neronov2010evidence, Taylor:2011bn, Takahashi:2011ac, Vovk:2011aa, durrer2013cosmological, acciari2023lower, aharonian2023constraints}. Furthermore, electromagnetic cascades, influenced by the IGMF, are referred to as "gamma-ray halos" and appear to be bow-tie-shaped structures surrounding point sources in the gamma-ray sky \cite{Broderick:2016akd}, although such phenomena have not yet been observed.\\ 
\indent However, alternative hypotheses, including collective plasma effects, can elucidate the phenomenon of cascade emissions. The interaction between the blazar-induced pair beam and the background intergalactic plasma can lead to the growth of plasma instabilities. These instabilities can be either electrostatic or electromagnetic in nature \cite{bret2005electromagnetic}. The collective beam-plasma instabilities can contribute to the energy loss compared to IC cooling. However, the efficiency of the energy loss due to plasma instability is still under debate as the studies by \cite{Broderick:2011av, Miniati:2012ge, schlickeiser2012plasma, schlickeiser2013plasma, Sironi:2013qfa, vafin2018electrostatic, AlvesBatista:2019ipr} only considered the linear evolution to estimate the energy-loss due to instabilities and excluded the non-linear feedback on the beam evolution. \cite{Castro:2024ooo} conducted a parametric study on the energy-loss length due to plasma instability and the instability power index for the real blazar source 1ES 0229+200. They estimated that the secondary electron pairs lose approximately 1\% of their energy over the typical interaction length for IC scattering based on their best-fit scenario. On the other hand, \cite{Alawashra:2024fsz} studied the non-linear feedback of electrostatic oblique instability for the same source using a Fokker-Planck equation coupled with the linear wave equation without any contribution of the IGMF. They found that the instability broadens the beam and leads to a minimal energy transfer from the beam to the plasma waves. In this paper, we investigate the instability growth in the linear phase and the feedback of instabilities in the non-linear phase for a non-Maxwellian beam (because of the warm and non-monoenergetic nature of pair-beams generated in blazars) using particle-in-cell (PIC) simulations, improving upon previous studies.\\
\indent Several approaches have been proposed to mimic this phenomenon in the laboratory, but the primary challenge is the generation of a neutral pair beam. Nevertheless, it is important to emphasize that most traditional beam optics components are not designed to manage beams with both electrons and positrons. Earlier experimental studies at high-intensity laser facilities have reported the ratio of positrons to electrons ($N_{e^{+}}/N_{e^{-}}$), including OMEGA-EP 2014 ($ \sim 10\%$) \cite{chen2014magnetic}, Orion/OMEGA-EP 2015 ($ \sim 10\%$) \cite{chen2015scaling}, Texas-Petawatt (PW) Laser ($ \sim 50\%$) \cite{liang2015high}, ASTRA-GEMINI Laser system ($96\%$) \cite{sarri2015generation, hooker2006astra}, OMEGA-EP 2021 ($ \sim 100\%$) \cite{peebles2021magnetically}, and HiRadMat ($97\%$) \cite{Arrowsmith:2023lhw}. The laboratory experiments provide an excellent opportunity to compare their outcomes with numerical results. Our research contributes to the understanding of how realistically scaled parameters can be selected in order to replicate astrophysical jets within a laboratory setting and also the complete beam evolution for a real astrophysical scenario.\\
\indent We consider a laboratory-based setup to investigate the evolution of instability with density contrast in the linear and non-linear phases using numerical simulations. For an astrophysical pair beam, \cite{rafighi2017plasma} established the criteria for setting up a physically relevant simulation: 
\begin{enumerate}[i]
    \item The kinetic energy density ratio of the beam to the background, $\epsilon = \alpha (\gamma - 1) m_{e}c^{2}/k_{B}T_{bg}$  should be less than unity, where $\gamma$ is the Lorentz factor and $T_{bg}$ defines the temperature of the background plasma,
    \item The electrostatic instability growth should dominate over electromagnetic instabilities, as the beam has a momentum anisotropy and a large longitudinal momentum spread. 
\end{enumerate}
In our study, we fix the kinetic energy density ratio to a value smaller than 1 that satisfies first condition and conduct an in-depth investigation into the density contrast range at which electrostatic instability is dominating over electromagnetic instability that satisfy the second criterion for a broad (or warm) non-Maxwellian beam under laboratory conditions. 
Subsequently, we extrapolate the nonlinear regime from laboratory to astrophysical scales to investigate the beam energy loss due to instabilities, as well as the impact of nonlinear feedback on the angular broadening of beams originating from a realistic blazar source. We assume that instability is the dominant mechanism in comparison to the IC cooling for astrophysical pair beams in order to examine the feedback of beam-plasma instability on the pair beam.\\
\indent The paper is structured as follows: In Section \ref{sec:beam-plasma-instability}, we outline the linear growth rates of beam-plasma instabilities, then in Section \ref{sec:beam-distribution}, we describe the configuration of a broad (or warm) non-Maxwellian beam distribution function, which is similar to the properties of an astrophysical pair beam. In Section \ref{sec:pic-sim}, we present the results of the PIC simulation of a beam-plasma system that can be replicated in a laboratory environment. Section \ref{sec:budget} presents the fractional beam energy loss due to instability. Section \ref{sec:angular-broadening} presents the angular broadening in the non-linear regime, which is interpreted as the feedback of instability. Next, in Section \ref{sec:implications-for-blazars}, we estimate the fractional energy loss of the beam and transverse broadening of the beam due to instability feedback for 1ES 0229+200-like sources. Finally, Section \ref{sec:conclusion} provides the conclusions of our findings.

\section{\label{sec:beam-plasma-instability}Beam Plasma Instability}
Our study focuses on the unstable dynamics of an ultra-relativistic neutral beam composed of electrons and positrons following the same momentum distribution. The beam propagates in a neutral background plasma consisting of electrons with no bulk velocity and immobile protons without having an external magnetic field. The system is characterized using two key parameters: the bulk Lorentz factor $\gamma$ of the pair beam and the density ratio $\alpha = n_{b0}/n_{bg}$ representing the peak beam density ($n_{b0}$) relative to the background plasma density ($n_{bg}$). Given our focus on the behavior of ultra-relativistic dilute beams, we can reasonably consider that $\alpha \ll 1$ and $\gamma \gg 1$. 
The current filamentation instability (Cfi) is an electromagnetic instability characterized by the excitation of both electric and magnetic field modes, with the unstable modes oriented perpendicular to the beam momentum. In contrast, the oblique instability (Obl) represents an electrostatic mode in which the unstable modes are oriented obliquely to the beam momentum. 
Using the definition of plasma frequency, $\omega_{p}=(4\pi n_{bg}e^{2}/m_{e}c^{2})^{1/2}$, the maximum theoretical linear growth rates of the dominant modes for an ultra-relativistic ($v_{b}\sim c$) beam can be described as follows \cite{bret2010multidimensional},
\begin{eqnarray}
    &\delta_{\text{Cfi,theory}} = \left(\frac{\alpha}{\gamma}\right)^{1/2} \omega_{p},\\
    &\delta_{\text{Obl,theory}} = \frac{\sqrt{3}}{2^{4/3}}\left(\frac{\alpha}{\gamma}\right)^{1/3} \omega_{p}.
\end{eqnarray}
We investigate the dominant growth rates of these instabilities for a warm non-Maxwellian beam, considering a set of physical parameters relevant to laboratory conditions, and likely extrapolate to model the astrophysical scenario.

\section{Realistic Pair Beam Distribution Function}\label{sec:beam-distribution}
Previous studies \cite{Broderick:2016akd, Broderick:2011av, Miniati:2012ge, schlickeiser2012plasma, schlickeiser2013plasma, Sironi:2013qfa, AlvesBatista:2019ipr, beck2023evolution, kempf2016energy, rafighi2017plasma, Yan:2018pca, Chang:2013hia, Chang:2014cta, Supsar_2014, Chang_2016, Shalaby:2017kpr, Tiede:2017xng, Shalaby:2018jja, Shalaby:2020fnm} on astrophysical plasma scenarios, specifically, collisionless space plasmas, are focused on Maxwellian non-monoenergetic beams having a small energy spread. This does not accurately represent real scenarios due to the broad nature of astrophysical pair distributions. Although Maxwellian beams are straightforward to generate in simulations, a more effective approach would involve superimposing two or more relativistic Cauchy (Breit-Wigner) distribution beams to better replicate relatively broad (or warm) non-Maxwellian beams \cite{Arrowsmith:2023lhw}. To compare the plasma properties in these two cases, we employ the concept of plasma screening length, or Debye screening, which refers to the ability of a plasma to shield or screen out electric fields over short distances. We evaluate the plasma screening length ($\lambda_{\text{scr}}$) for both Maxwellian and Cauchy distribution functions by applying the limit for the static field ($\omega/k \to 0$) of the dielectric tensor. Thus, the longitudinal component of the dielectric tensor reaches a finite value. As the screening length is applicable primarily at large distances, we use the long-wavelength limit, setting $k\to 0$. Given that the axisymmetry, $k_{\parallel}=k_{x}$ and $k_{\bot}=(k_{y}^{2} + k_{z}^{2})^{1/2}$ can be assumed for a neutral pair beam without affecting generality. Under the electrostatic approximation (i.e., $\mathbf{k}\times\mathbf{E}=\mathbf{0}$), the dielectric tensor can be expressed as follows \cite{bret2005electromagnetic, bret2010exact, bret2010multidimensional, ghosh2022light},
\begin{equation}
    \varepsilon = 1 + \sum_{s} \frac{m_{s}\omega_{p,s}^{2}}{k^{2}}\int \frac{\mathbf{k}\cdot\mathbf{\nabla_{p}}f_{s}(p)}{\omega - \mathbf{k}\cdot\mathbf{v}}d^{3}\mathbf{p},
    \label{eqn:die}
\end{equation}
where $p$ denotes the normalized momentum, $f_{s}$ represents the normalized distribution function, $\omega_{p,s}=(4\pi n_{s}q_{s}^{2}/m_{s}c^{2})^{1/2}$ is the plasma frequency, and $n_{s}$ defines the number density for each species $s$. For a beam with arbitrary velocity $\mathbf{v}$ and wave vector $\mathbf{k}$, the filamentation mode corresponds to $\mathbf{k} \cdot \hat{\mathbf{v}} = 0$ and oblique modes corresponds to $\mathbf{k} \cdot \hat{\mathbf{v}} = \omega \cos\theta$. For unstable modes, the associated plasma waves satisfy the resonance condition $\omega - \mathbf{k} \cdot \mathbf{v} = 0$. 

The plasma screening length can be evaluated as \cite{lifshitz1981physical, silin1960electromagnetic},
\begin{equation}
    \begin{split}
    \lambda_{\text{scr}}^{-2} &= \lim_{k\to 0}\left\{\lim_{\omega/k \to 0}k^{2}\left(\varepsilon -1 \right)\right\}\\&
    = -4\pi \sum_{s} \frac{\omega_{p,s}^{2}}{c^{2}}\int_{0}^{\infty} \gamma p \frac{\partial f_{s}\left(p\right)}{\partial p} dp,
    \label{eqn:screening}
    \end{split}
\end{equation}
For a relativistic scenario, a simplistic normalized Maxwellian beam can be described as follows,
\begin{equation}
    f_{s}\left(p\right) = \frac{1}{4 \pi \sigma_{\parallel,0} K_{2}\left(1/\sigma_{\parallel,0}\right)}\exp{\left(-\frac{p}{\sigma_{\parallel,0}}\right)},
    \label{eqn:maxwell-mono}
\end{equation}
where $K_{2}(1/\sigma_{\parallel,0})$ is the modified Bessel function of the second kind. Since the pair beam distribution functions are non-thermal, we define $\sigma_{\parallel,0}$ as the initial longitudinal momentum spread. In astrophysical jets, beam particles have an enhanced high-energy tail compared to a Maxwellian distribution. Considering a simple case, the normalized suprathermal beam distribution can be written as follows,
\begin{equation}
    f_{s}\left(p\right) = \frac{1}{\pi^{2}}\left(1 + \frac{p^{2}}{\sigma_{\parallel,0}^{2}}\right)^{-2},
    \label{eqn:cauchy-mono}
\end{equation}
Therefore, the plasma screening length for the Maxwellian beam described by Eq.~(\ref{eqn:maxwell-mono}) is calculated as $\lambda_{\text{scr}}\simeq \sigma_{\parallel,0}^{1/2}c\cdot\omega_{p,s}^{-1}$. For the suprathermal Cauchy beam described by Eq.~(\ref{eqn:cauchy-mono}), the expression becomes  $\lambda_{\text{scr}}\simeq 0.88 \sigma_{\parallel,0}^{1/2}c\cdot\omega_{p,s}^{-1}$. The plasma screening lengths for Maxwellian and Cauchy distributions are approximately the same, indicating that in both cases, the fundamental behavior is similar, differing only in the high-energy tail.
The pair beam distribution for astrophysical jets is non-thermal, featuring high-energy tails and broad nature, in contrast to a simplistic Maxwell-J\"{u}ttner distribution. A suprathermal kappa distribution function exhibits high-energy tails: 
\begin{equation}
\begin{split}
    f(\textbf{p}; \mu, \sigma_{\parallel, \perp}, \kappa) \propto &\frac{\Gamma\left(\kappa + 1\right)}{2\pi\kappa^{3/2}\Gamma\left(3/2 \right)\Gamma\left(\kappa - 1/2 \right)}\times \\& \left[1 +\gamma^{2}\frac{\left(p_{x} - \mu\right)^{2}}{2\kappa\sigma_{\parallel}^{2}} + \gamma^{2}\frac{p_{y}^{2} + p_{z}^{2}}{2\kappa\sigma_{\perp}^{2}}\right]^{-\left(\kappa + 1\right)},
\end{split}
\end{equation}
For large values of $\kappa$, the distribution function approaches a Gaussian. However, since our focus is on high-energy power-law tails, we use $\kappa=1$ to model a non-Maxwellian distribution (which is close to the Cauchy distribution function) appropriate for astrophysical contexts. 
To better capture this, we investigate the evolution of an ultra-relativistic beam with a realistic Cauchy distribution in our simulation,
\begin{equation}
    f(\textbf{p}; \mu, \sigma_{\parallel, \perp}) \propto \left[1 +\gamma^{2}\frac{\left(p_{x} - \mu\right)^{2}}{2\sigma_{\parallel,0}^{2}} + \gamma^{2}\frac{p_{y}^{2} + p_{z}^{2}}{2\sigma_{\perp,0}^{2}}\right]^{-2}, 
	\label{eqn:dist-fn}
\end{equation}
In the present work, we use a system of ''plasma units'' where the fundamental constants, $\omega_{p} = c = k_{B} = 1$. In this study, the distribution function of electron-positron ($e^{+}e^{-}$) pairs produced is non-thermal.  To replicate an astrophysical pair beam in the laboratory, we must account for it producing electron-positron pairs with an opening angle of approximately $\theta_{0} \sim \gamma^{-1}$. Considering a longitudinally warm or broad and transversely kinematically cold beam, the initial transverse momentum spread can be scaled as,
\begin{equation}
    \sigma_{\bot,0} = p \sin(\theta_{0}) = \gamma m_{e} \sin(\gamma^{-1})\sim m_{e}.
\end{equation}
This suggests that the highly relativistic particle will exhibit a wide range of longitudinal momenta, and a transverse momentum spread approximately equal to the electron mass. This configuration closely resembles the characteristics of an astrophysical pair beam.\\
\begin{figure}[ht]\centering
\includegraphics[width=0.45\textwidth]{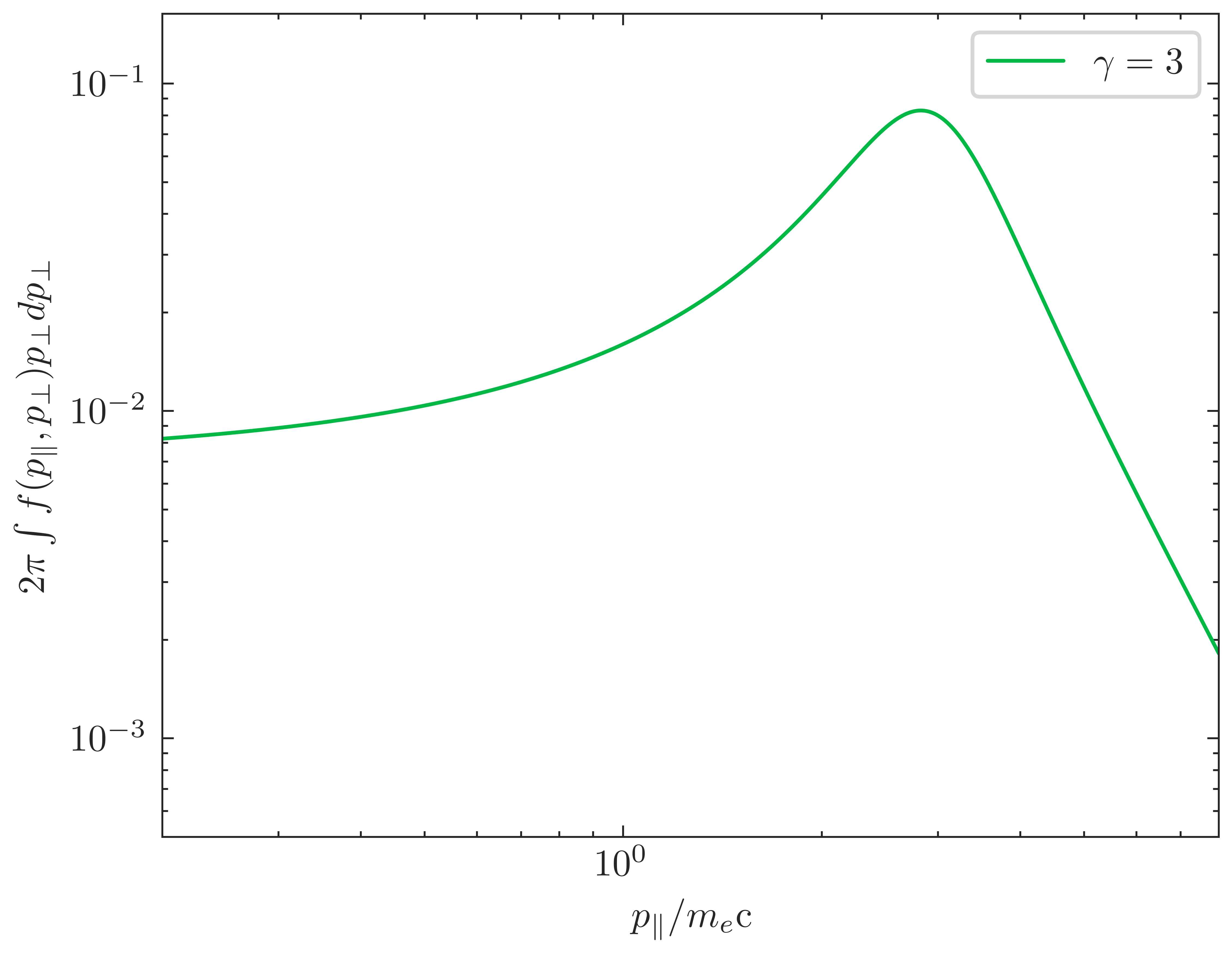}
\caption{The distribution function described by Eq. (\ref{eqn:dist-fn}) for a broad beam with $\gamma = 3$ and range of $\alpha$ values, is used in PIC simulations. The initial longitudinal momentum spread, $\sigma_{\parallel, 0}=1.0$ MeV, the initial transverse momentum spread, $\sigma_{\perp, 0}= 0.5$ MeV, and a mean longitudinal momentum, $\mu=0.511(\gamma^{2}-1)^{1/2}~\text{MeV}$ is defined for the beam.}
\label{fig:dist}
\end{figure}
Fig.~\ref{fig:dist} shows the pair beam distribution following Eq.~(\ref{eqn:dist-fn}), integrated across the transverse momentum. The following distribution function perfectly resembles the non-Maxwellian nature of the beam at high energies.

\section{PIC Simulations}\label{sec:pic-sim}

\subsection{Simulation set up}\label{subsec:sim-set-up}
In order to model the propagation of a warm beam through a background plasma in a two-dimensional Cartesian configuration, we employ EPOCH-2D PIC simulation code \cite{Ridgers:2015ikp}. In the initial condition setup, the length of simulation box of the longitudinal direction is defined by $L_{\parallel,box} = 120 c\cdot{\omega_{p}^{-1}}$ and the transversal direction is represented by $L_{\perp,box}=120 c\cdot{\omega_{p}^{-1}}$. The number of cells along x- and y-direction is $N_{x} = N_{y} = 880$ and each cell contains $N_{p} = 200$ particles per species. The total number of particles per species is $N_{tot} = 1.5488\times 10^{8}$. We use four types of particles: beam electrons and beam positrons exhibiting relatively broad Cauchy momentum distribution, background plasma consisting of electrons with no bulk velocity, and background protons, which are immobile due to their larger mass.  
The initial beam density profile is given by $n_b = n_{b0} [1 + (y/R_{y,0})^{2}]^{-2}$ where $R_{y,0} = 110 c \cdot \omega_{p}^{-1}$ denotes the initial rms beam transverse spatial width, the initial peak beam density $n_{b0}= \alpha n_{bg}$, and the background particles have a density of $n_{bg} = 10^{16}$ cm$^{-3}$ for different values of $\alpha$ in the simulations. In laboratory experiments, the transverse spatial width of relativistic pair beams is typically on the order of mm to cm \cite{sarri2015generation, arrowsmith2021generating, Arrowsmith:2023lhw}. In our study, where the simulations are performed on laboratory scales, the skin depth $c/\omega_{p} \sim 0.53$ mm, and the transverse beam width we have used is greater than the skin depth, which is consistent in the laboratory regime. In our series of simulations, we explore $\alpha$ values of 0.0005, 0.005, and 0.05 while maintaining a fixed Lorentz factor, $\gamma = 3$ for all cases. Although these $\alpha$ values are significantly higher than those relevant to real TeV Blazars, the chosen range within the laboratory regime provides ample scope for reliable extrapolation to lower values. We employ periodic boundary conditions in the longitudinal direction and open boundary conditions in the transverse direction to simulate a finite-width beam. Periodic boundary conditions ensure that fields and particles arriving at one side of the simulation box reappear at the opposite side, but for the open boundary system, particles simply transmit through the boundary and leave the system. The beam width is smaller than the transverse size of the simulation box, $L_{y,box}$, and does not reach the boundaries. Therefore, the use of open (or absorbing) boundary conditions does not significantly influence beam-edge effects, as it provides the physical consistency of the system and closely resembles a beam with a finite width. In contrast, a periodic boundary represents an infinitely extended system, and a reflecting boundary reflects particles at the boundary, potentially causing artificial field enhancements within the simulation box. Neither is appropriate for a finite-width beam configuration. Since we don't want to lose the background plasma particles, we apply thermal boundary conditions for the background in both x and y directions. In the case of a thermal boundary condition, whenever a particle exits the simulation domain, it is replaced by an incoming particle with the same temperature. Simulations are conducted up to the total time $4500 \omega_{p}^{-1}$ to capture the linear, the non-linear growth phase, and the saturation region. The criteria for a physically relevant configuration of the beam-plasma system are that the kinetic energy density ratio must satisfy the condition $\epsilon <1$. In accordance with the previous section, the warm neutral pair beam is characterized by maintaining an initial longitudinal momentum spread, $\sigma_{\parallel, 0}=1.0$ MeV with a mean longitudinal momentum, $\mu=0.511(\gamma^{2}-1)^{1/2}~\text{MeV}$ and an initial transverse momentum spread, $\sigma_{\perp, 0}= 0.5$ MeV for each sub-beams. This indicates that in the background plasma rest frame, the beam is initially transversely cold and longitudinally warm as the initial angular spread, $\theta_{0} \equiv \sigma_{\bot, 0}/\sigma_{\parallel, 0} = 0.5$, which is less than 1. A comprehensive overview of the simulation parameters is presented in Table \ref{tab:ex-table}, and Table \ref{tab:ex-table1} outlines the sub-beam parameters.
\begin{table}[ht]
	\centering
	\caption{The summary of the PIC Simulation configuration for the simulation run.}
	\label{tab:ex-table}
	\begin{tabular}{l*{6}{c}r}\hline
        Parameters & Value\\
        \hline
Number of Dimensions & 2 ($x$ beam dir., $y$ trans.)\\
$L_{x,box} (L_{\parallel,box})$ & 120 $c\cdot{\omega_{p}^{-1}}$ \\
$L_{y,box} (L_{\perp,box})$ & 120 $c\cdot{\omega_{p}^{-1}}$  \\
No. of cells along x-direction ($N_x$) & 880 \\
No. of cells along y-direction ($N_y$) & 880 \\
No. of particles per cell $N_p$ (per species) & 200 \\
Total no. of particles ($N_{tot}$) & $1.5488\times 10^{8}$ \\
Timestep $\Delta t$ & 0.95 CFL-Criterion \\
Maxwell Solver & Yee (2nd order) \\
Finite difference scheme & 6th order \\
Particle Pusher & Higuera \& Cray\\
Particle Shape Function & Third Order B-Spline\\
Current Filtering & 5-fold (1-2-3-4 steps) \\
$n_{bg}$ & $10^{16}$ cm$^{-3}$ \\
$T_{bg}$ & $\alpha(\left\langle\gamma\right\rangle - 1)600$ keV \\
Background particles & Immobile Protons \& $e^{-}$\\
Beam particles & $e^{-}$ \& $e^{+}$ \\
Distribution Function & Eq.~(\ref{eqn:dist-fn}) \\
$\epsilon$ & 0.85\\
Total time $T$ & 4500 $\omega_{p}^{-1}$\\
\hline\hline
Boundary conditions & \\
\hline
For the beam and fields (along x) & periodic\\
For the beam and fields (along y) & open\\
For the background (along x \& y) & thermal\\
\hline
\end{tabular}
\end{table}

\begin{table}[ht]
	\centering
	\caption{The overview of the beam parameters for the Cauchy beam distribution function.}
	\label{tab:ex-table1}
	\begin{tabular}{l*{6}{c}r}\hline
Parameters & Beam\\
\hline
$\gamma$ & 3\\
$\mu$ (MeV) & $m_{e}(\gamma^{2}-1)^{1/2}$ \\
$\sigma_{\parallel, 0}$ (MeV) (alias $\Delta p_{x,0}$) & 1.0\\
$\sigma_{\perp, 0}$ (MeV) (alias $\Delta p_{y,0}$) & 0.5\\
\hline
	\end{tabular}
\end{table}
\subsection{Growth of plasma instabilities and evolution of fields}\label{subsec:inst-growth}
The primary numerical measurable quantity that we can access in a simulation run is the growth of the fields. The initial noise due to the thermal fluctuation of the background plasma can influence the evolution of the fields. The initial noise of the system is inversely proportional to $\epsilon$, i.e., Initial noise $\propto 1/(N_{tot}\epsilon)$, where $N_{tot}$ is the total number of simulation particles \cite{beck2023numerical}. Increasing the background plasma temperature results in a decrease of the parameter $\epsilon$. Therefore, we increase $N_{tot}$ to offset the significant noise caused by high background plasma temperatures. Throughout these simulations, we maintain a very low initial noise level. We employ a Yee Maxwell field solver with 6th-order field interpolation and a 3rd-order B-spline shape function (yielding a 5th-order weighting) for placing particles on the grid. The default multiplying factor for this field solver is set to 0.95 of the Courant–Friedrichs–Lewy (CFL) criterion on time steps. A particle pusher is implemented following the method described by \cite{higuera2017structure}. To reduce numerical noise, we apply a 5-fold current smoothing, following the approach outlined by \cite{vay2014modeling}.
\begin{figure}[ht]
\centering
\includegraphics[width=0.43\textwidth]{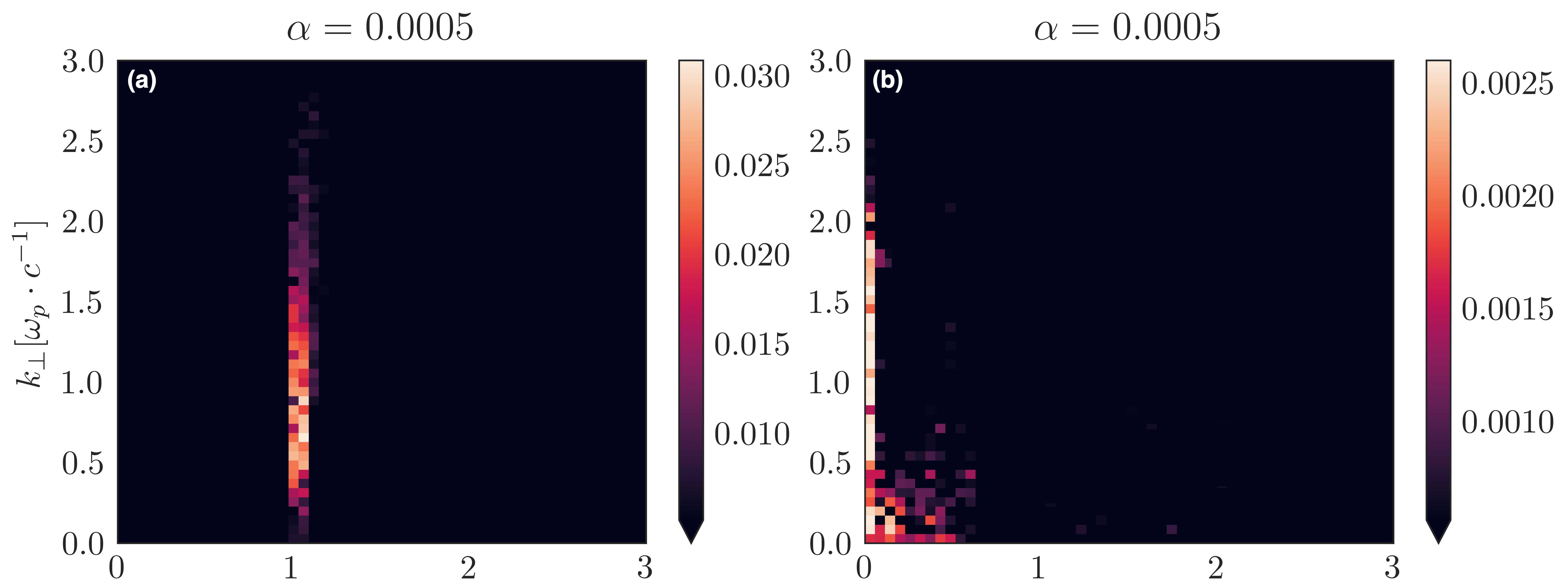}
\includegraphics[width=0.43\textwidth]{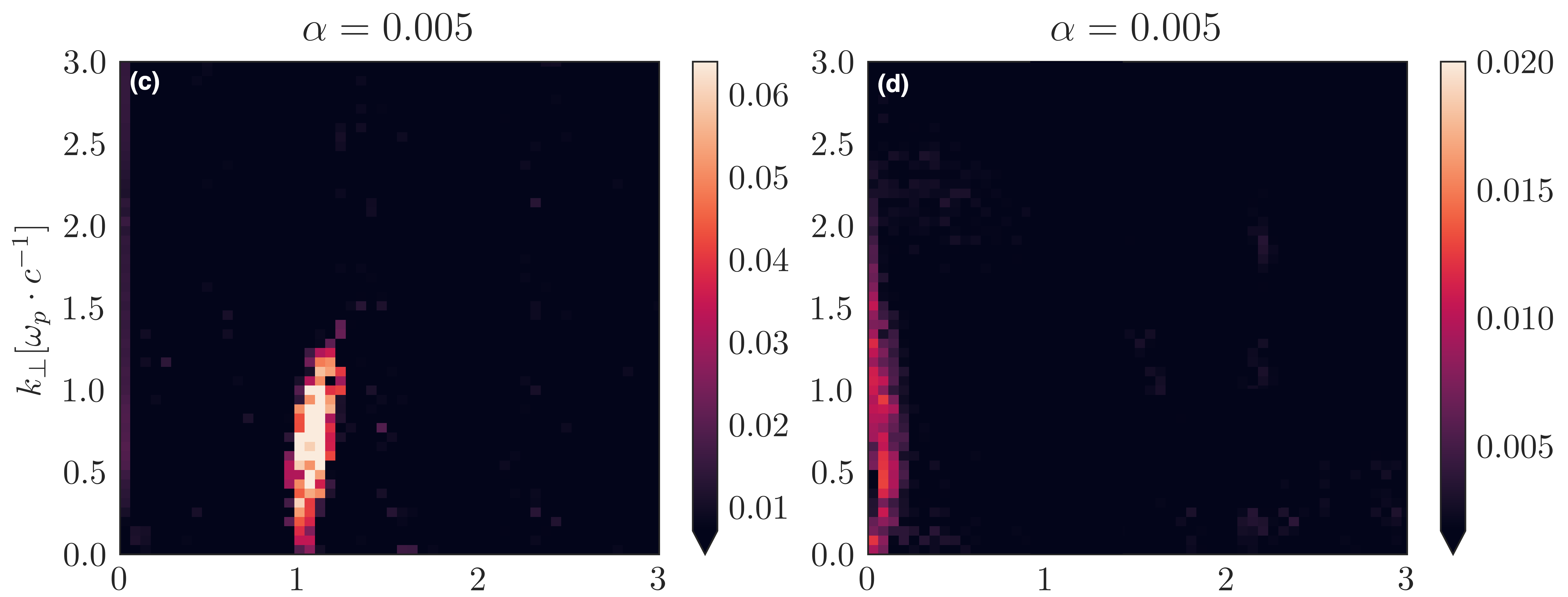}
\includegraphics[width=0.43\textwidth]{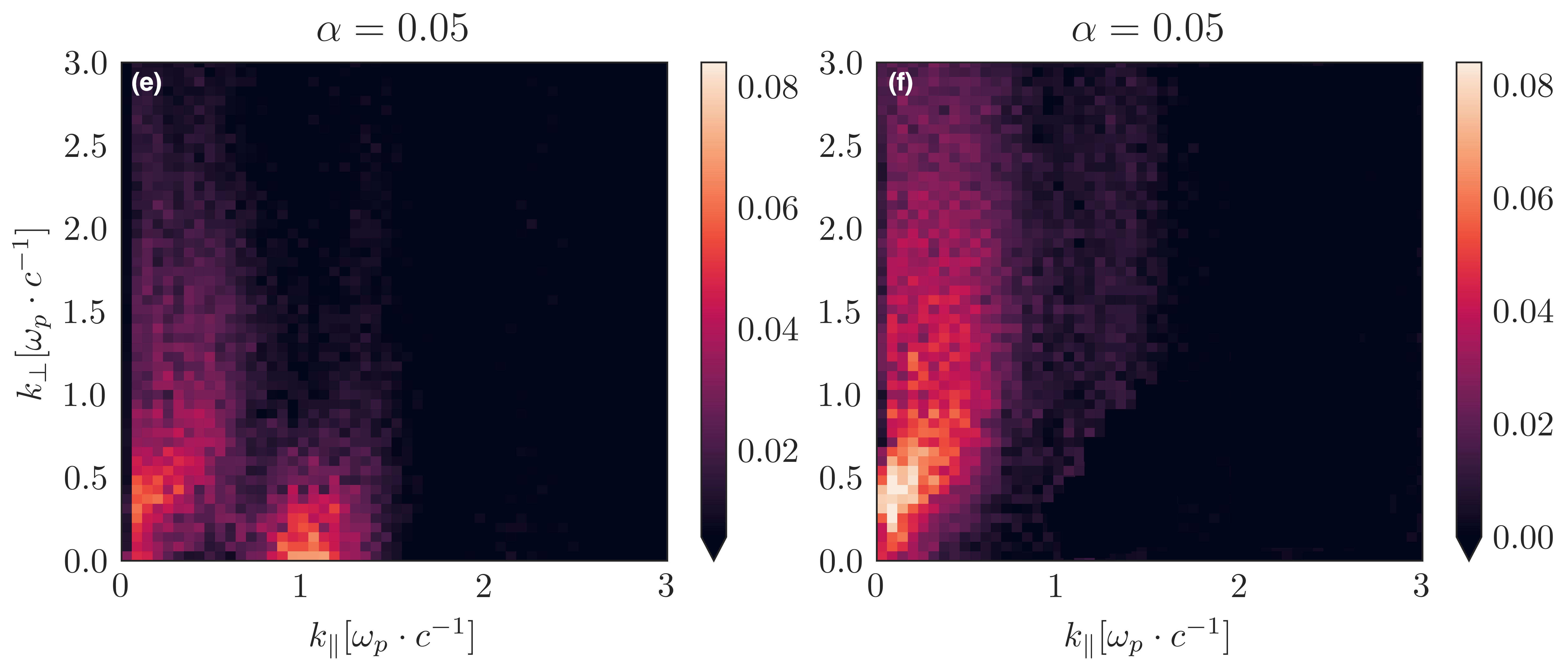}
\caption{Two-dimensional snapshots of the growth rate of instability in $\mathbf{k}$ space. Figs. (a), (c), and (e) along the \textbf{left} column show the growth rates of electrostatic instability, which are obtained from Fourier transformation of $E_x$ and $E_y$ fields. Figs. (b), (d), and (f) along the \textbf{right} column show the growth of B-modes, which are obtained from Fourier transformation of the $B_x$ and $B_z$ fields. The color scale defines the growth rate which is plotted as a function of $\delta(k_{\parallel},k_{\bot})[\text{in units of }\omega_{p}]$ for $\alpha = 0.0005$ at $t \omega_{p}\sim 670$, $\alpha = 0.005$ at $t \omega_{p}\sim 310$, and $\alpha = 0.05$ at $t \omega_{p}\sim 170$. With increasing $\alpha$, the electrostatic and magnetic modes become comparable, and the system gradually transitions to a regime where electromagnetic modes dominate.}
\label{fig:growth-sim1}
\end{figure}
\begin{figure}[ht]
\centering
\includegraphics[width=0.47\textwidth]{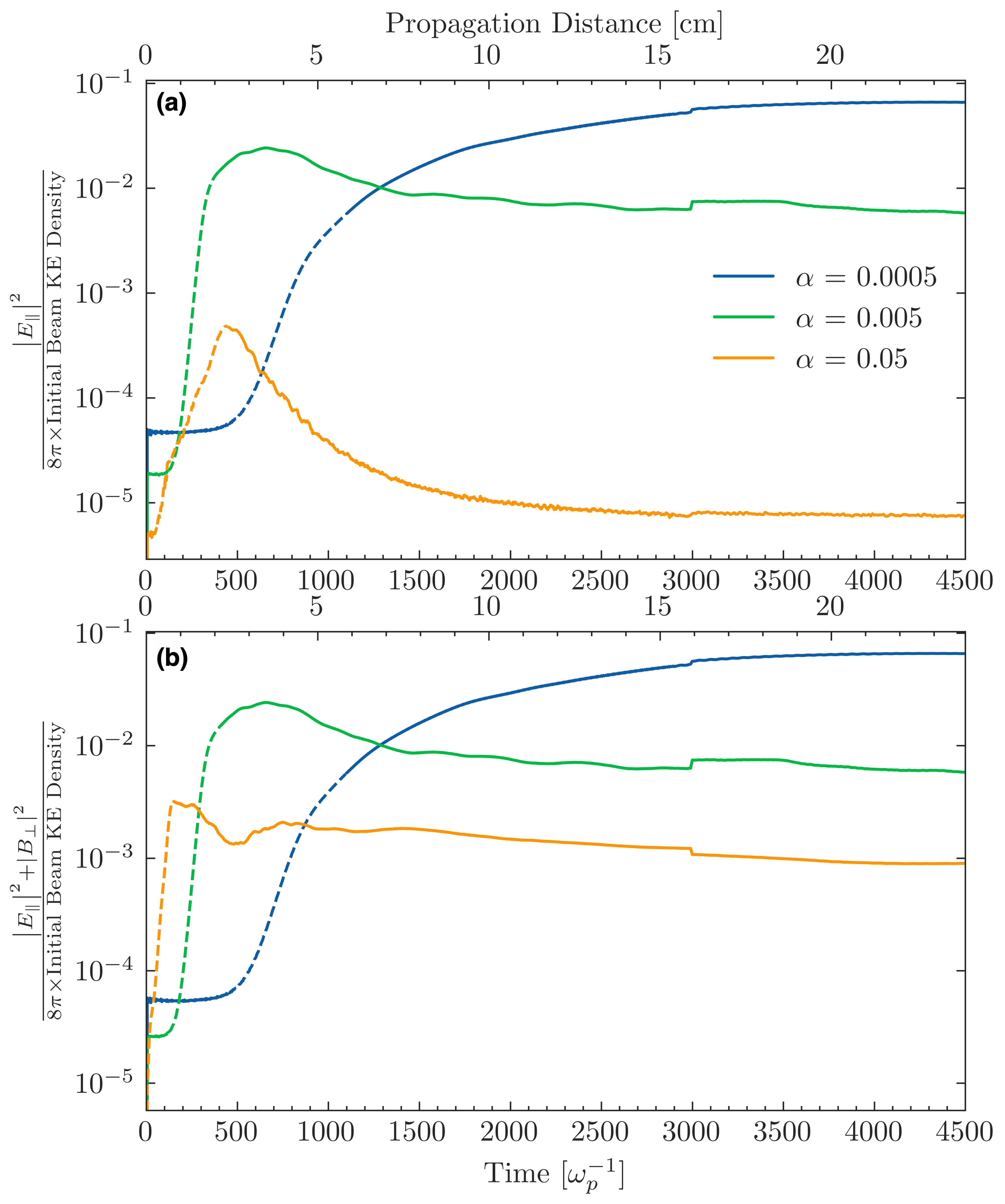}
\caption{The fraction of beam kinetic energy converted into the (a) longitudinal electric field, (b) electromagnetic field across all modes is shown with respect to time on the bottom x-axis and beam propagation distance on the top x-axis. The beam parameters outlined in Table \ref{tab:ex-table1} are applied in every simulation for different density contrast $\alpha$. The dashed lines represent the linear growth phase for each $\alpha$, with lower values of $\alpha$ leading to a greater dominance of the electric field modes. Whereas, for $\alpha=0.05$, electromagnetic modes begin to dominate.}
\label{fig:e-parallel} 
\end{figure} 
\begin{figure}[ht]
\centering
\includegraphics[width=0.47\textwidth]{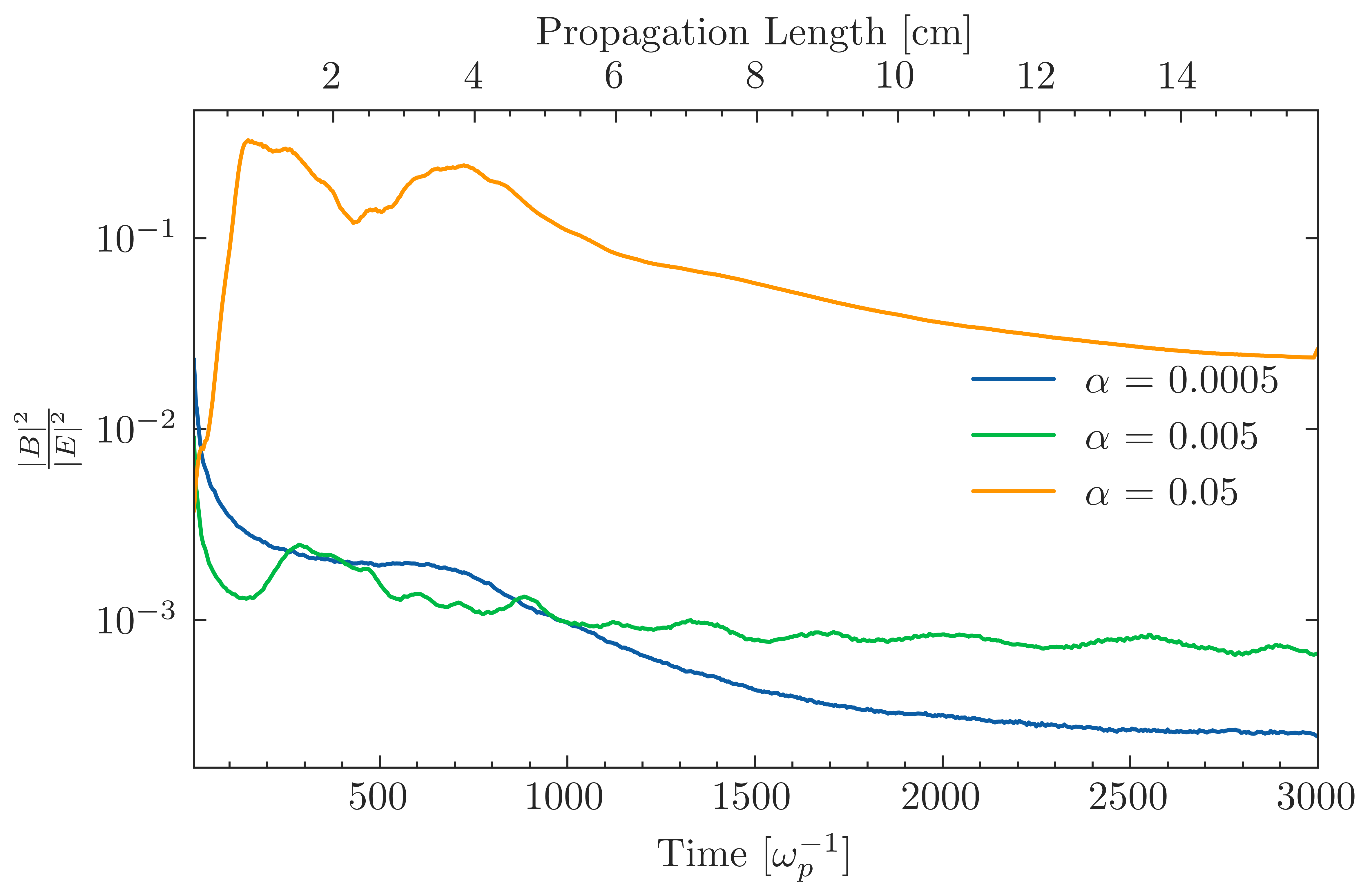}
\caption{The comparison of magnetic field to electric field strength across all modes is shown as a function of time (bottom x-axis) and beam propagation distance (top x-axis). The beam parameters specified in Table \ref{tab:ex-table1} are used for all simulations with varying density contrast $\alpha$, similar to Fig.~\ref{fig:e-parallel}. As $\alpha$ decreases, the magnetic field strength becomes weaker than the electric field strength.}
\label{fig:b-e-ratio} 
\end{figure}
The wave number is expressed as having a component parallel to the direction of propagation of the beam, denoted as $k_{\parallel} = k_{x}$ and the component perpendicular to the beam propagation is given by $k_{\bot} = ({k_{y}^{2} + k_{z}^{2}})^{1/2}$. In Figs.~\ref{fig:growth-sim1}a, \ref{fig:growth-sim1}c, and \ref{fig:growth-sim1}e, we present the growth rates of electrostatic instability, which is plotted against wavevectors for different $\alpha$ values with beam parameters specified in Table~\ref{tab:ex-table1}. The Figs.~\ref{fig:growth-sim1}b, \ref{fig:growth-sim1}d, and \ref{fig:growth-sim1}f show the growth rates of electromagnetic instability, which is plotted against wavevectors for the same set of $\alpha$ values. The growth rate is defined as a function of $\delta(k_\parallel, k_\perp)$, extracted by performing a Fourier transform of the $E_x$ and $E_y$ fields, followed by computing the Fourier amplitude for each mode. The resonant electric field mode is the fastest growing mode for $\alpha = 0.0005$ and $0.005$. The maximum oblique growth is observed around $k_{\parallel}\sim c\cdot \omega_{p}^{-1}$. In contrast, the magnetic field modes appear comparatively weak for $\alpha = 0.0005$ and $0.005$. The electric and magnetic field modes become comparable for $\alpha =0.05$. The maximum filamentation growth is observed around $k_{\parallel}\sim 0$. In Fig.~\ref{fig:growth-sim1}e, resonant modes around $k_{\parallel}\sim 0$ are also observed, as the $E_y$ component is also associated with the electromagnetic filamentation mode.
The transition between the dominant current filamentation modes to oblique modes is immediately observed while decreasing $\alpha$ from $0.05$ to $0.005$. Fig.~\ref{fig:e-parallel}a illustrates the time evolution of the longitudinal electric fields for different values of $\alpha$. As $\alpha$ increases, a significantly smaller fraction of the beam kinetic energy is converted into longitudinal electric fields, indicating that the oblique mode faints as the beam becomes relatively denser. The Fig.~\ref{fig:e-parallel}b shows the beam is losing less than 1\% of its kinetic energy during the linear growth phase, contributing to the growth of electromagnetic fields. We find that the beam energy loss rate reduces after the linear growth phase ends, and the system continues to evolve non-linearly. By the end of saturation, the diluted beam lost approximately 7\% of its kinetic energy, contributing to the growth of electromagnetic fields.
Fig.~\ref{fig:b-e-ratio} shows that as $\alpha$ decreases, the magnetic field strength increases compared to the electric field. The magnetic field amplification is triggered between $\alpha=0.005$ and $0.05$ since electromagnetic instability starts to dominate. 
Consequently, for ultra-relativistic dilute beams (i.e., $\alpha \ll 1$), the condition of growth rate, $\delta \ll 1$ is satisfied for the electrostatic modes. For instance, when $\alpha = 0.005$ and Lorentz factor $\gamma = 3$, the theoretical electrostatic growth rate is calculated as $\delta_{Obl,theo} \sim 0.071\omega_{p}$. The maximum electrostatic growth rate measured in the simulation is approximately $\delta_{Obl,sim}\sim 0.064\omega_{p}$. This fulfills the consistency of the simulation with the analytical results.

\subsection{Interplay between magnetic field and beam momentum}\label{subsec:mag-field-growth}
\begin{figure}\centering
\includegraphics[width=0.47\textwidth]{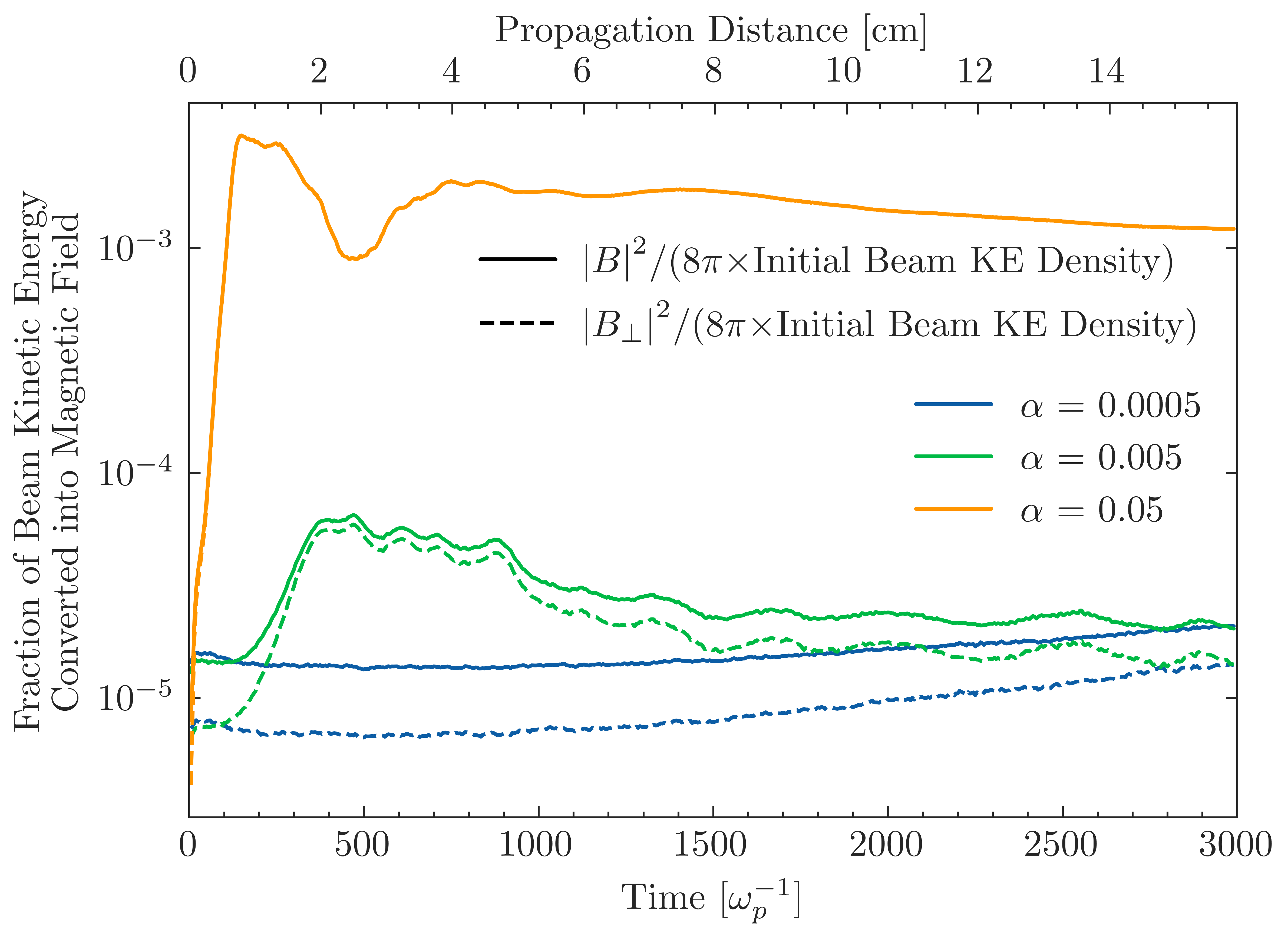}
\caption{The fraction of beam kinetic energy transferred into the magnetic field across all modes is shown with respect to time (bottom x-axis) and the distance the beam travels during this time (top x-axis). For every simulation with varying density contrast $\alpha$, the beam parameters provided in Table \ref{tab:ex-table1} are applied similarly as Fig.~\ref{fig:e-parallel} and \ref{fig:b-e-ratio}. The solid lines represent the total magnetic field, and the dashed lines depict only the transverse component of the magnetic field.}
\label{fig:b-field}
\end{figure}

\begin{figure}[ht]
\centering
\includegraphics[width=0.42\textwidth]{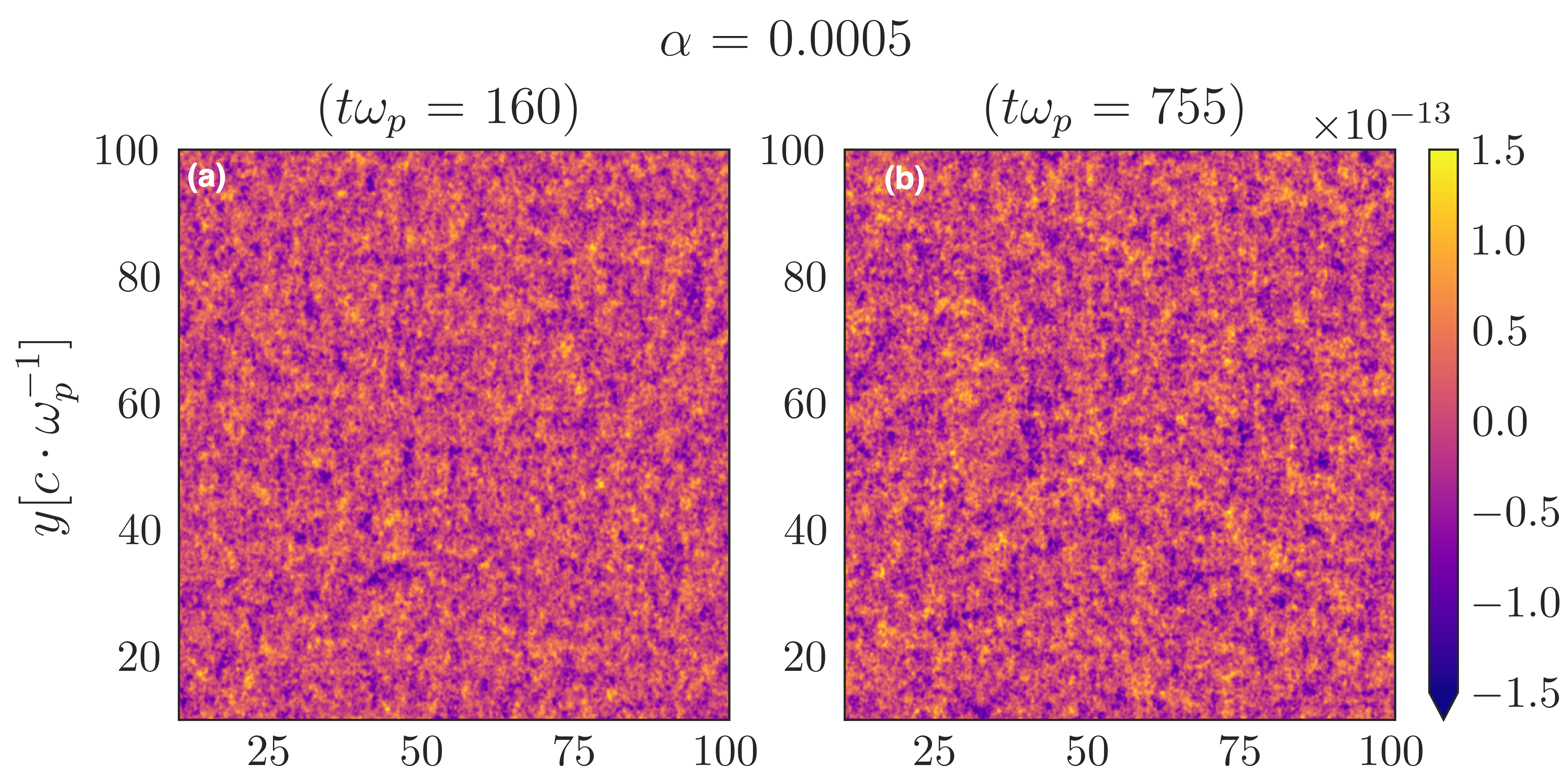}
\includegraphics[width=0.42\textwidth]{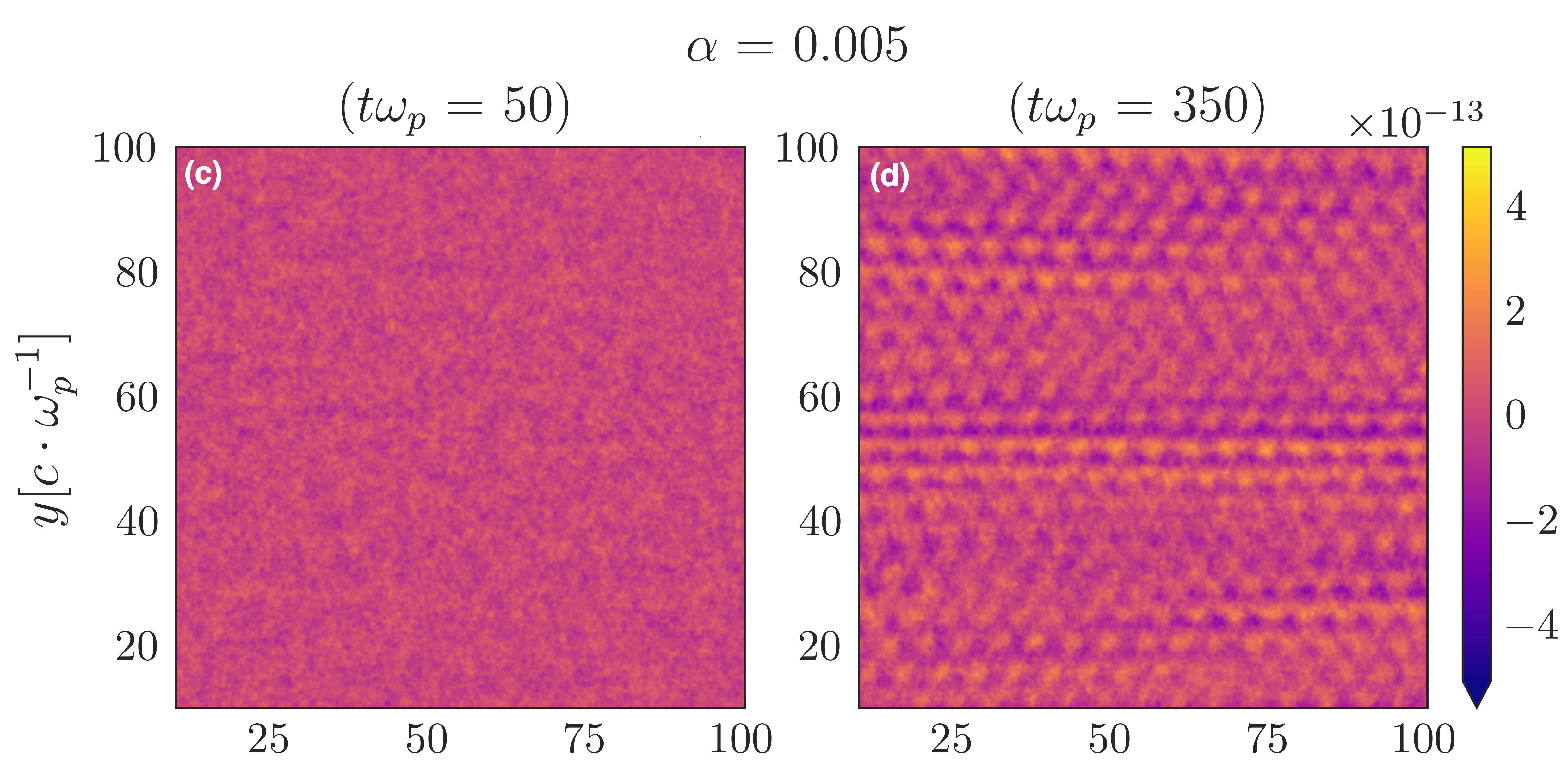}
\includegraphics[width=0.42\textwidth]{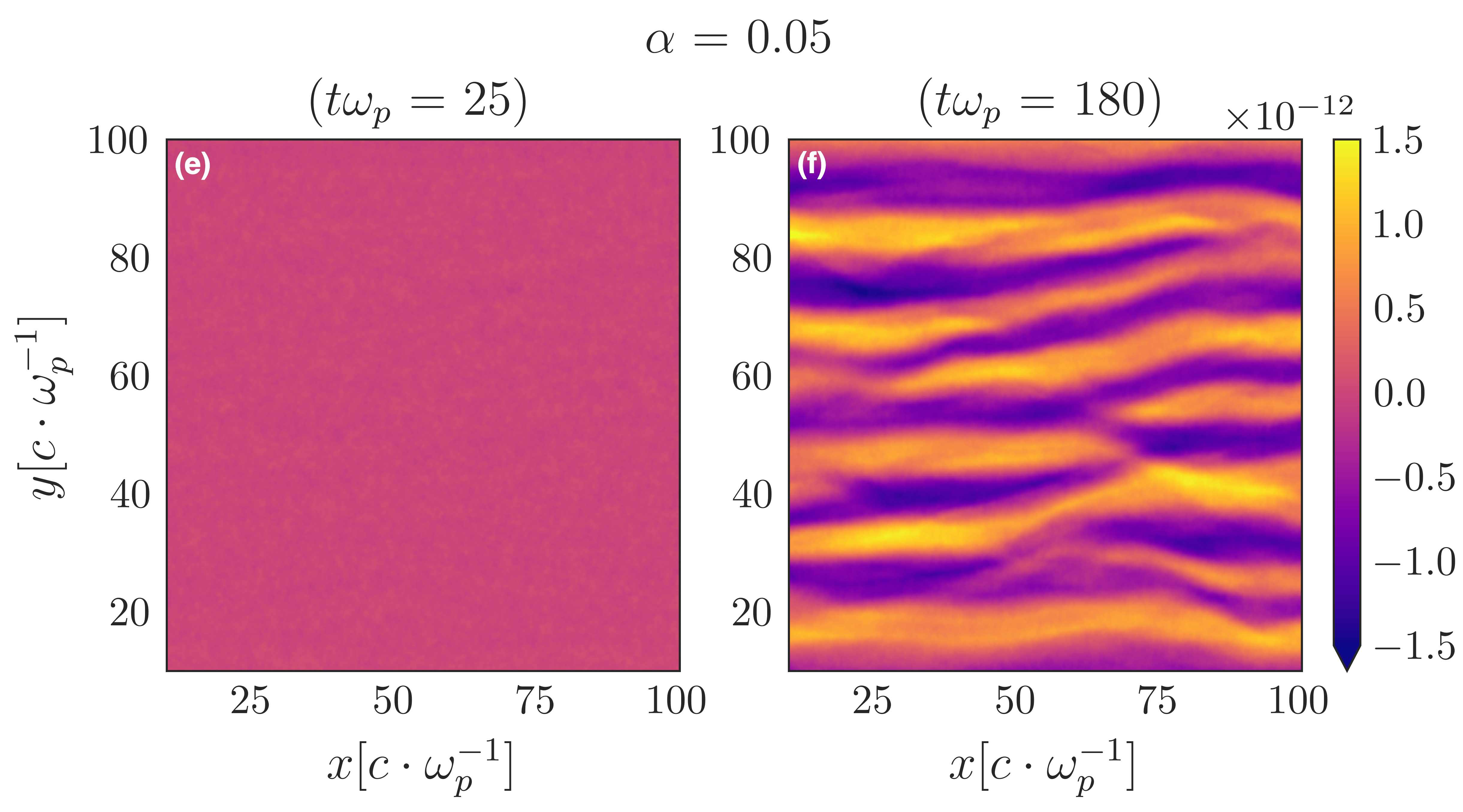}
\caption{The evolution of the transverse magnetic field with varying density contrast $\alpha$. The color scale represents the fraction of beam kinetic energy converted into each mode of the associated transverse magnetic field, i.e., $(B_{\bot})_{k}/\sqrt{\text{Beam KE Density}}$. The Figs. (a), (c), and (e) along the \textbf{left} column illustrate the state before the linear growth phase of the instability sets in, whereas Figs. (b), (d), and (f) along the \textbf{right} column show the evolution of the transverse magnetic field at the instability growth phase.}
\label{fig:filament}
\end{figure}

\begin{figure}[ht]
\centering
\includegraphics[width=0.45\textwidth]{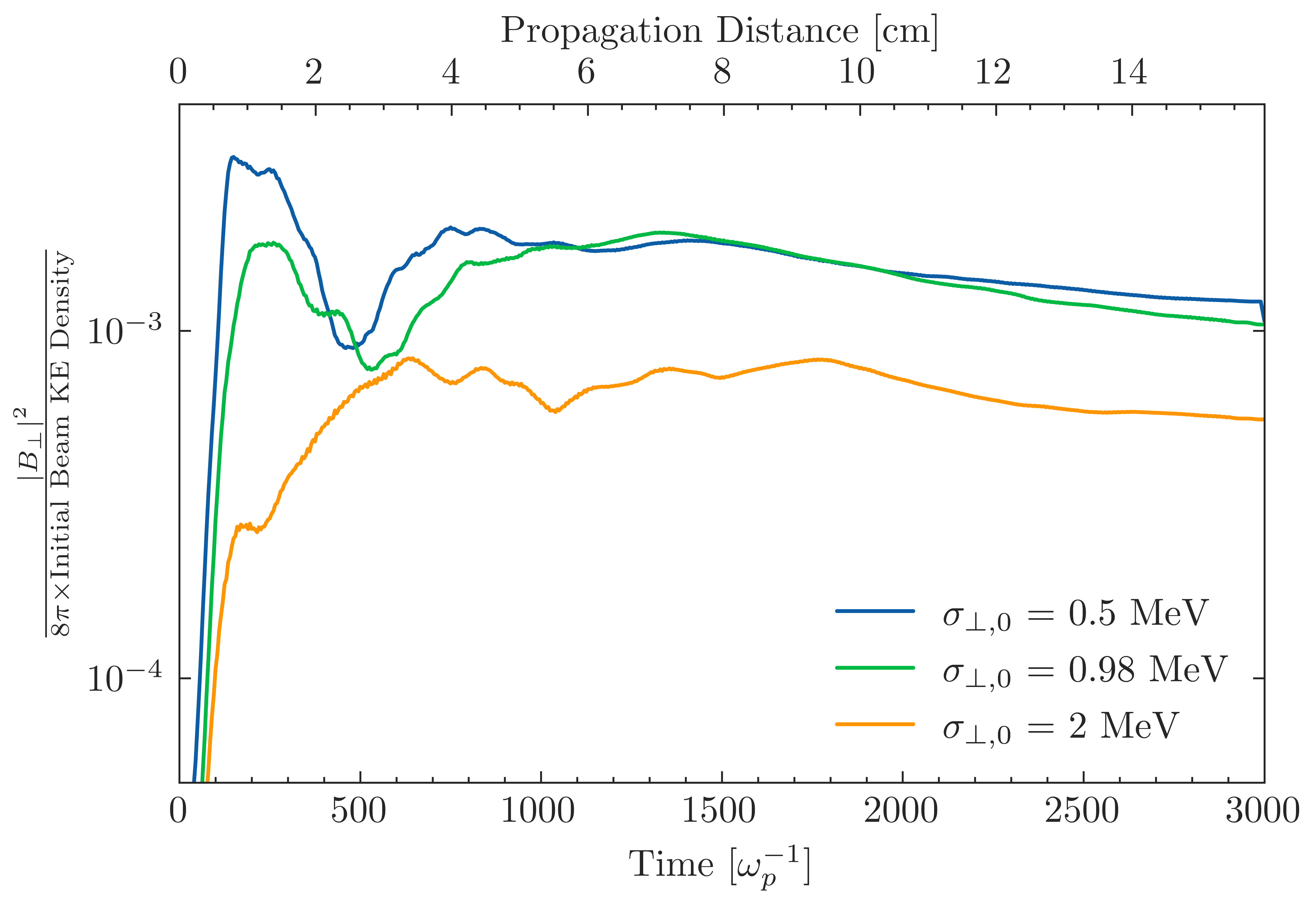}
\caption{The fraction of beam kinetic energy transferred into the transverse magnetic field across all modes with respect to time is represented on the bottom x-axis, and the propagation distance is shown on the top x-axis. In every simulation, the density contrast is fixed at $\alpha = 0.05$ and $\sigma_{\parallel,0} = 1.0~\text{MeV}$, while the initial transverse momentum of the beam is varied. The beam parameters listed in Table~\ref{tab:ex-table1} are used. As the initial transverse momentum increases or the beam becomes colder, both the current filamentation instability and the secondary growth are diminished.}
\label{fig:b-p}
\end{figure}

\begin{figure*}[t!]
\centering
\includegraphics[width=0.9\textwidth]{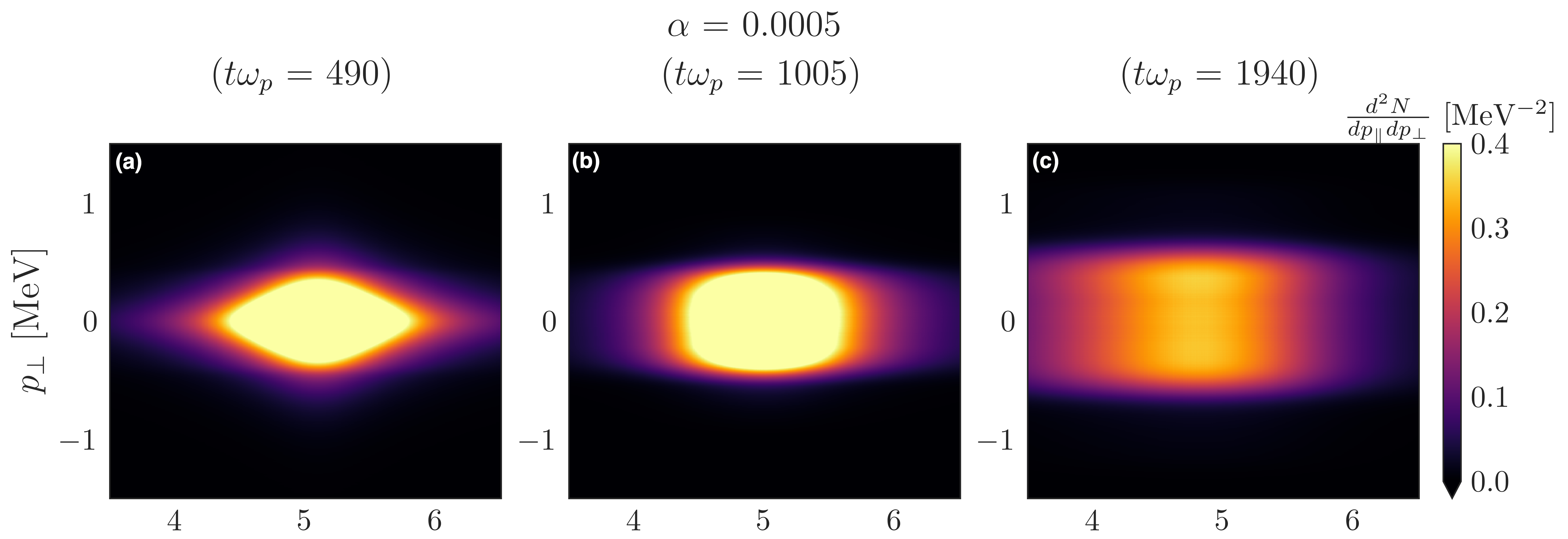}
\includegraphics[width=0.9\textwidth]{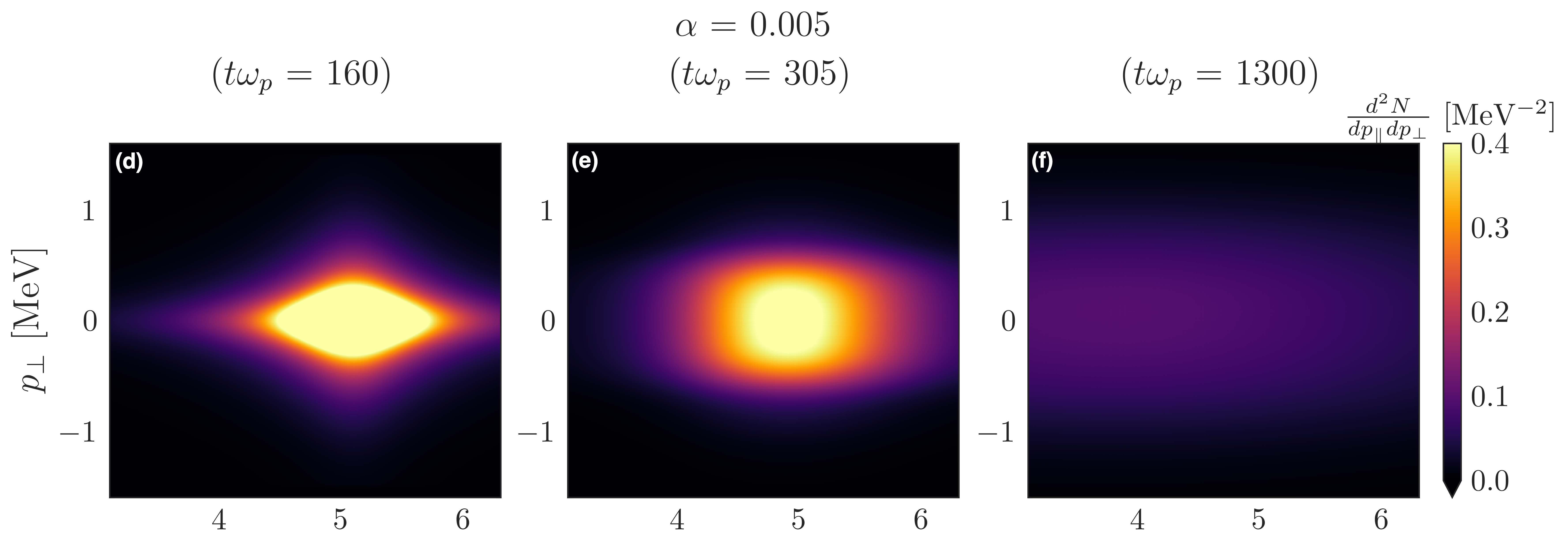}
\includegraphics[width=0.9\textwidth]{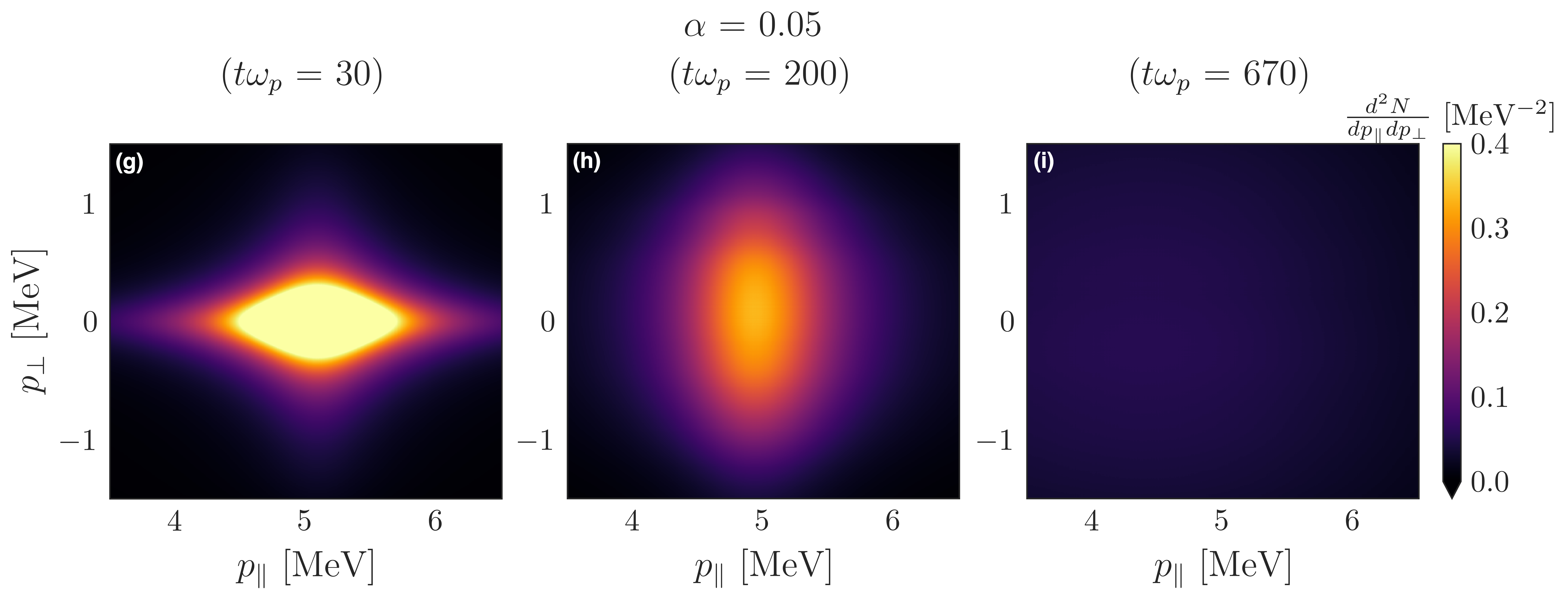}
\caption{Momentum distribution in two dimensions for different stages of instability growth for different values of $\alpha$. The Figs. (a), (d), and (g) on the \textbf{left row:} before starting the linear growth, Figs. (b), (e), and (h) on the \textbf{middle row:} during the linear growth phase, and Figs. (c), (f), and (i) on the \textbf{right row:} at the end of the non-linear regime (starts to saturate). The color scale represents the beam distribution function in momentum space and has a unit of MeV$^{-2}$. The different timestamps for instability growths depend on the dominant instability modes according to Table \ref{tab:data}.}
\label{fig:mom}
\end{figure*}

When a neutral pair beam propagates through a plasma, small perturbations can arise, leading to the spatial separation of electrons and positrons in the beam, resulting in the formation of localized currents. The separation can be on the scale of the skin depth of the beam. The localized currents can create filaments, leading to the generation of magnetic fields. Fig.~\ref{fig:b-field} shows the fraction of beam kinetic energy converted into the magnetic field for different $\alpha$ values. For increasing values of $\alpha$, the magnetic field strength increases because the electron and positron filaments start to separate spatially, producing self-generated localized currents \cite{sarri2015generation}. The electromagnetic instability grows until the wavelength of unstable modes is comparable to the Larmor radius of the particles in the self-generated magnetic field, causing them to become trapped in the magnetic fields \cite{ davidson1972nonlinear2, achterberg2007weibel}. Consequently, an electromagnetic counterpart emerges despite the overall beam remaining neutral. Accordingly, at higher density contrast, magnetic fields are generated and the filamentation instability develops because of the transverse magnetic pressure linked to the filaments, causing the total magnetic field to predominantly align in the transverse direction $\left(\lvert B\rvert \sim \lvert B_{\bot}\rvert\right)$. However, after the linear growth phase, there is a secular growth of filamentation instability before saturation (which starts at $t\sim 510\omega_{p}^{-1}$, as shown in Fig.~\ref{fig:b-field}), particularly for $\alpha = 0.05$. This is due to the formation of a small cavity, which is more clearly observed in the spatial structure of the magnetic field in a 3D simulation \cite{Peterson:2021yit, Peterson:2022cyw}. As the magnetic pressure within these unscreened cavities causes them to expand, more current is exposed, leading to secular growth. The secondary growth will eventually saturate as the net current in the cavity decreases.
Fig.~\ref{fig:filament} depicts the evolution of the spatial structure of the transverse magnetic field at the initial state and during the primary linear growth phase. As $\alpha$ increases, the filamentation instability gives rise to the development of distinct transverse magnetic filament structures. The phenomenon of generation of current filamentation has been studied explicitly using both analytical and semi-analytical methods by \cite{weibel1959spontaneously, yoon1987exact, sakai2000magnetic, silva2002role, jaroschek2005ultrarelativistic, Sironi:2013qfa, Groselj:2024dnv}. Fig.~\ref{fig:b-p} illustrates the fraction of beam kinetic energy transferred to the transverse magnetic field for different initial transverse beam momentum spreads. As the initial transverse beam momentum increases, with $\sigma_{\parallel,0} = 1.0~\text{MeV}$ fixed, the induced transverse magnetic field decreases, suggesting a decrease in transverse current filamentation instability in the linear growth regime.
However, while a warm beam can drive current filamentation instability during the linear growth phase, it also enhances the conditions for secondary filamentation growth in the non-linear regime. Fig.~\ref{fig:mom} shows the two-dimensional momentum distribution at different simulation timestamps for various values of $\alpha$. Due to the different growth rates associated with varying $\alpha$ values, the linear growth phase begins and ends at different times. At the start of the linear growth phase, the beam is focused, maintaining its stability without significant perturbations affecting the longitudinal beam momentum ($p_{\parallel}$) and the transverse beam momentum ($p_{\bot}$). As the instability growth progresses, the beam spreads energetically both in the transverse and longitudinal direction, resulting in an overall broadening of its distribution. The transversal broadening in the non-linear phase is reduced as the beam gets diluted because the electromagnetic modes become suppressed. As previously explained, in the filamentation-instability-driven scenario, a secondary filamentation instability develops during the nonlinear phase. This leads to a non-linear feedback that significantly causes strong transverse momentum broadening.\\ 
Table \ref{tab:data} summarizes the dominant instabilities that are responsible for momentum broadening in the non-linear phase across the range of $\alpha$ values considered in our simulation. The beam with $\alpha = 0.005$ is the threshold on a laboratory scale where the non-linear feedback of electrostatic instability is observed, leading to the energetic broadening of the beam over time.
\begin{table}
	\centering
	\caption{Summary of the dominant instabilities in the linear growth phase that are responsible for momentum broadening in non-linear regions for different $\alpha$ runs.}
 \label{tab:data}
	\begin{tabular}{l*{6}{c}r} \hline
$\theta_{0}~[\text{rad}]$ & $\gamma$ & $\alpha$ & Dominant instability\\
\hline
{} & {} & 0.0005 & Oblique\\
0.5 & 3 & 0.005 & Oblique\\
{} & {} & 0.05 & Transverse current filamentation\\
\hline
	\end{tabular}
\end{table}

\section{Beam energy-loss}\label{sec:budget}
\begin{figure}[h]
\centering
\includegraphics[width=0.38\textwidth]{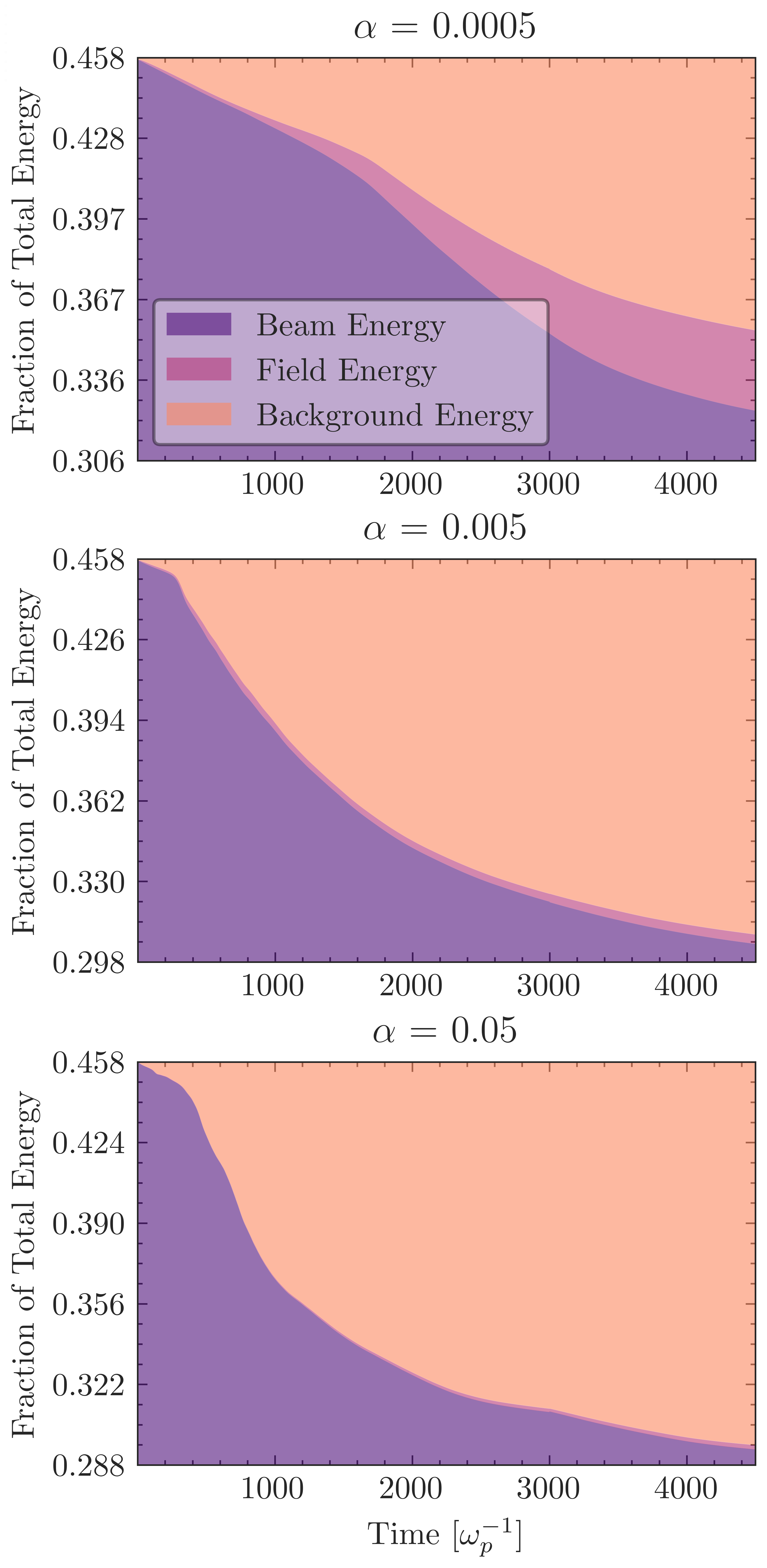}
\caption{
The allocation of the total energy of the system between the beam kinetic energy, the total electromagnetic field energy, and the energy of the background medium is presented over time with different density contrast $\alpha$. Each simulation run uses beam parameters as detailed in Table \ref{tab:ex-table1}.}
\label{fig:budget-fig}
\end{figure}

\begin{figure}[h]
\centering
\includegraphics[width=0.45\textwidth]{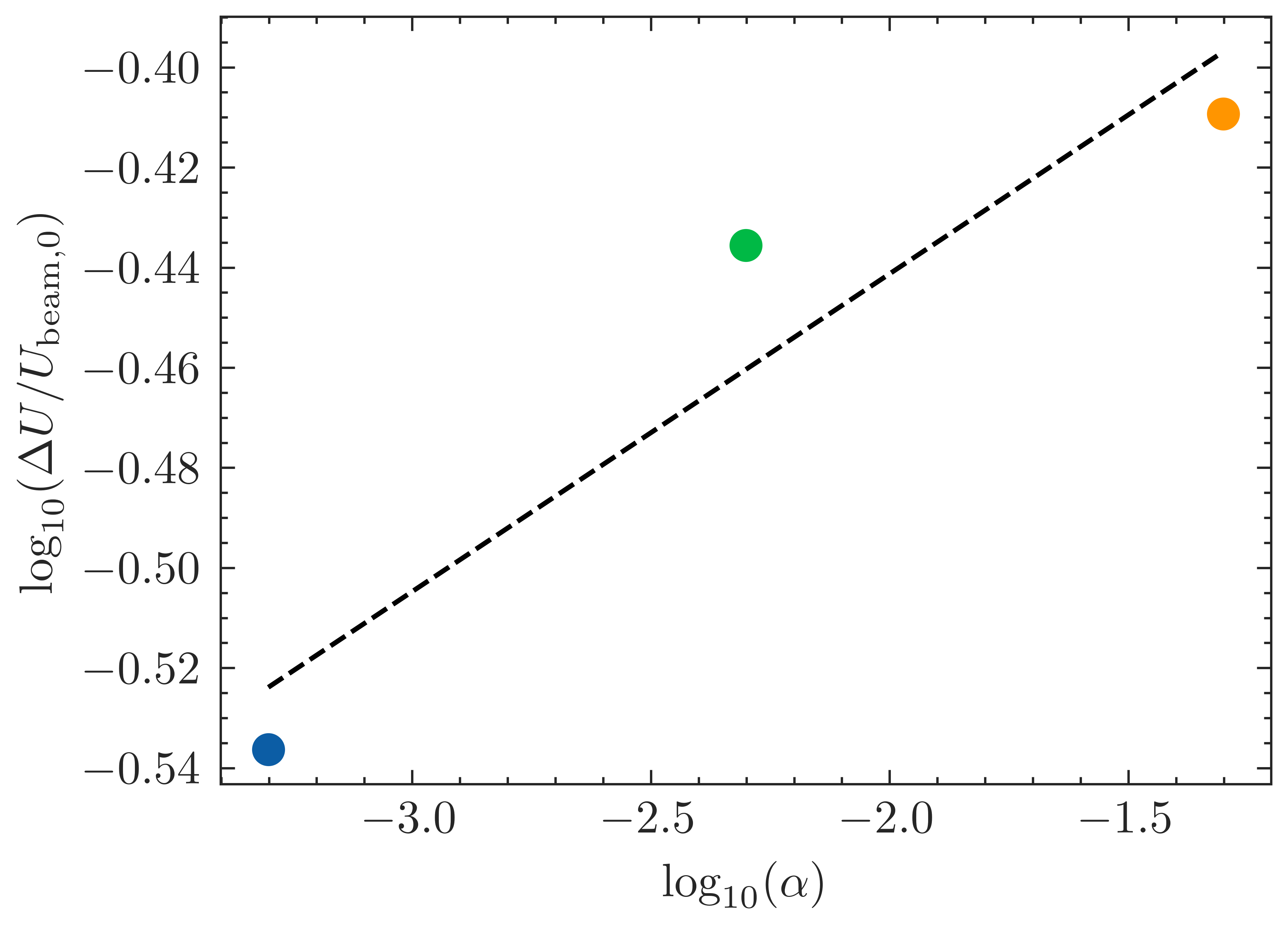}
\caption{The power-law scaling of the fractional beam energy loss, $\Delta U/ U_{\text{beam},0}$ with $\alpha$ at the saturation. The black dashed line represents the fitted line. The fitted power-law index for the $\alpha$ scaling is 0.07.}
\label{fig:energy-loss-scale}
\end{figure}

We have studied the quantitative estimation of the proportion of total energy of the system allocated to beam kinetic energy, total field energy, and the energy of background particles. For instance, based on the parameters listed in Table \ref{tab:ex-table1}, Fig.~\ref{fig:budget-fig} indicates that initially, around 45.8\% of the total energy is allocated to the kinetic energy of the beam, while approximately 54.2\% is distributed to the energy of the background particles. This reflects the initial condition of our simulation setup, where the ratio of the beam to background kinetic energy density is $\epsilon \sim 0.85$. The growth of plasma instabilities can facilitate energy transfer from the beam to the background. The energy transfer continues even after the instability growth phase ends. For beams with $\alpha = 0.0005$, 0.005, and 0.05, the total beam energy loss relative to the initial beam energy is approximately 29\%, 36\%, and 38\%, respectively. Of the initial beam energy, the contribution to field energy growth is about 7\% for $\alpha = 0.0005$, 1\% for $\alpha = 0.005$, and less than 1\% for $\alpha = 0.05$. We define the fractional beam energy loss as $\Delta U/ U_{\text{beam},0}$, where $\Delta U$ represents the total energy lost by the beam and $U_{\text{beam},0}$ is the initial kinetic energy of the beam. This quantity serves as a measure of how much of the initial beam energy is dissipated into fields or background plasma. We obtain the power-law scaling of the fractional beam energy loss with $\alpha$:
\begin{equation}
    \left(\frac{\Delta U}{U_{\text{beam},0}}\right) \sim 0.48\cdot~\alpha^{0.07},
\label{eqn:energy-loss}
\end{equation}
Fig.~\ref{fig:energy-loss-scale} shows the power-law scaling of the fractional beam energy loss with $\alpha$. This suggests that even at very small density contrasts, the energy loss remains non-negligible.

\section{Angular broadening of the beam}\label{sec:angular-broadening}
\begin{figure}[h]
\centering
\includegraphics[width=0.45\textwidth]{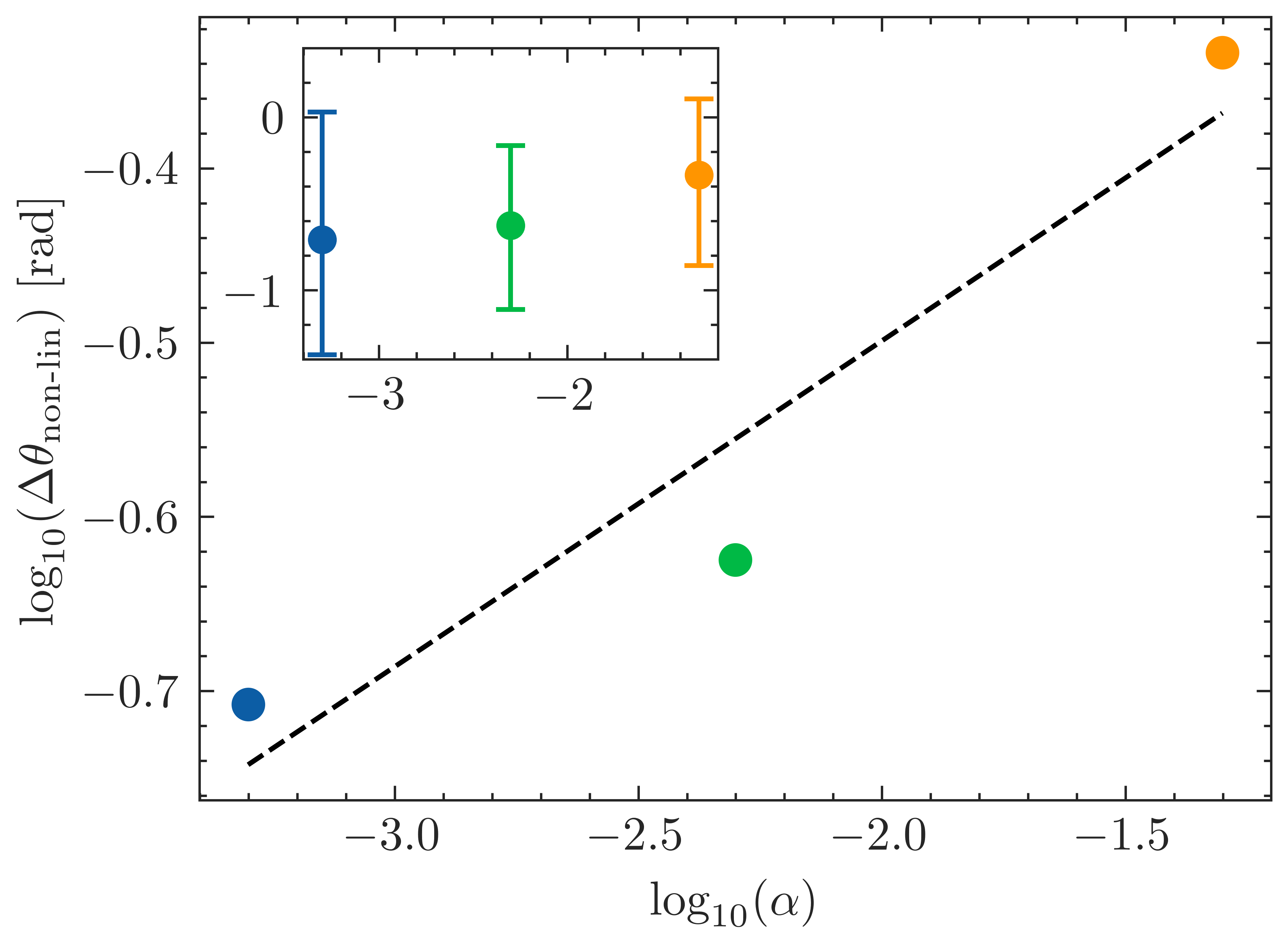}
\caption{The power-law scaling of $\Delta \theta_{\text{non-lin}}$ approximately at the end of non-linear phase with $\alpha$ is observed at $t\omega_{p}\simeq \{1940. 1300, 670\}$ respectively for $\alpha = \{0.0005, 0.005, 0.05\}$. This indicates the moment when the beam starts to enter the saturation phase. The black dashed line represents the fitted line. The embedded plot displays the error bars associated with the main plot (also shown in Table \ref{tab:data2}), with the errors calculated for $\Delta (t\omega_{p}) = \pm 10$.}
\label{fig:mom-with-time}
\end{figure}

Here we establish a power-law scaling relationship between the angular broadening ($\Delta \theta_{\text{non-lin}}$) due to non-linear instability feedback and $\alpha$, with the other parameters fixed as specified in Table \ref{tab:ex-table1}. The power-law scaling is estimated at the times when the beam has already formed its shape and enters into the saturation region. Fig.~\ref{fig:mom-with-time} illustrates the scaling of the $\Delta \theta_{\text{non-lin}}$ ($\equiv \Delta p_{\bot}/p_{\parallel}$) with $\alpha$ and the errorbars have been estimated with a deviation of $\Delta (t\omega_{p}) = \pm 10$, is shown in the inset plot. Table \ref{tab:data2} represents the data used to obtain the linear plot. The power-law scaling can be expressed as, 
\begin{equation}
    \left(\frac{\Delta \theta_{\text{non-lin}}}{1~\text{rad}}\right) \sim 0.75\cdot~\alpha^{0.19},
    \label{eqn:momentum-scaling}
\end{equation}
The dependence of $\Delta \theta_{\text{non-lin}}$ on $\alpha$ is relatively weak. 
\begin{table}
    \centering
    \caption{The overview of the $\Delta p_{\bot}$ data for different $\alpha$ values and the fitting parameters are obtained from the simulation. The errors are estimated for $\Delta (t\omega_{p}) = \pm 10$.}
    \label{tab:data2}
    {\renewcommand{\arraystretch}{1.2}
    \begin{tabular}{l*{6}{c}r}\hline
$\gamma$ & $\text{log}_{10}(\theta_{0})$ [rad] & $t\omega_{p}$ & $\alpha$ & $\text{log}_{10}(\Delta \theta_{\text{non-lin}})$ [rad]\\
\hline
{} & {} & 1940 & $0.0005$ & $-0.708^{+0.03}_{-1.37}$\\
$3$ & $-0.3$ & 1300 & $0.005$ & $-0.625^{-0.16}_{-1.11}$\\
{} & {} & 670 & $0.05$ & $-0.334^{+0.11}_{-0.86}$\\
		\hline
    \end{tabular}
    }
\end{table}

\section{Extrapolation for 1ES 0229+200-like sources}\label{sec:implications-for-blazars}
The total isotropic-equivalent luminosity ($\mathcal{L}$) of Blazar and photon energies influence the density of electron-positron pairs. The upper limit of the pair density can be derived assuming a balance between pair production and the IC cooling mechanism \cite{Broderick:2011av} as,
\begin{equation}
    n_{b} \simeq 1.9\cdot10^{-21}\text{cm}^{-3}\left(\frac{1+z}{2}\right)^{3\zeta-4}\left(\frac{\mathcal{L}}{10^{45}\text{erg/s}}\right)\left(\frac{\gamma}{10^{7}}\right),
\label{eqn:upper-limit-density}
\end{equation}
The parameter $\zeta = 4.5$ for $z < 1$ can be derived from the local star formation rate analysis \cite{Kneiske:2003tx}. As a benchmark scenario, we consider the blazar source 1ES 0229+200 located approximately at $z\approx 0.14$, fixing other parameters $n_{\text{IGM}} = 10^{-7}\text{cm}^{-3}$ and the fiducial value of $\mathcal{L} = 10^{45}\text{erg/s}$, consistent with \cite{AlvesBatista:2019ipr}. The density ratio can be expressed as:
\begin{equation}
    \alpha \simeq 9.1\cdot10^{-17}\left(\frac{\gamma}{10^{7}}\right),
\label{eqn:alpha-upper-limit}
\end{equation}
For a pair beam with a Lorentz factor of $10^7$, the corresponding density contrast is $\simeq 10^{-17}$. Using Eq.~(\ref{eqn:energy-loss}), the beam energy loss due to plasma instabilities can be estimated as,
\begin{equation}
    \left(\frac{\Delta U}{U_{\text{beam},0}}\right)\times 100\% \simeq 4.0\times\left(\frac{\gamma}{10^{7}}\right)^{0.07}\%,
\label{eqn:energy-loss-extrapolated}
\end{equation}
This implies that for a beam with $\gamma = 10^{7}$, the energy loss of the beam due to instabilities is approximately $4\%$.

The intrinsic opening angle of the pair beam with a Lorentz boost $10^7$ is determined by $\theta_{0} \sim 10^{-7}$ rad. The angular spread at the non-linear stage can be expressed from Eq.~(\ref{eqn:momentum-scaling}) as,
\begin{equation}
    \Delta \theta_{\text{non-lin}} = \frac{\Delta p_{\bot}}{p_{\parallel}}\Bigg|_{\text{non-lin}} \sim 6.7\cdot10^{-4}\left(\frac{\gamma}{10^{7}}\right)^{0.19}~\text{rad},
    \label{eqn:angle}
\end{equation}
The resulting angular spread due to the instability feedback for a beam with a Lorentz boost $10^{7}$ is nearly $6.7 \cdot 10^{-4}$ rad, which leads to a negligible angular broadening of the pair beam, considering the fact that the IC cooling rate is slower than the instability growth.

\section{Conclusions}\label{sec:conclusion}
We revisit the evolution of blazar-induced neutral pair beams under laboratory conditions using PIC simulations. The aim of our study is to estimate the density contrast range for which the beam can be considered dilute in the sense that electromagnetic instabilities are subdominant. In this parameter regime, the laboratory experiments can closely mimic the astrophysical case. The conclusions of this study can be summarized as follows:
\begin{itemize}
    \item In the absence of an external magnetic field, the dominant instability depends on the peak beam density. For a warm beam, at significantly higher beam density contrast ($\sim 0.05$), the beam is more prone to transverse current filamentation growth during the linear growth phase. The magnetic field amplification results from the combined effects of momentum anisotropy and beam-plasma instabilities. 
    \item Our results suggest that in laboratory experiments, a longitudinally broad (or warm) beam with an initial angular spread, $\theta_{0} = 0.5$ and $\alpha \leq 0.005$ can be classified as a ''dilute beam'', achieving a regime where oblique instability dominates dominates over the electromagnetic instabilities (as specified in Table \ref{tab:data}). In the non-linear regime, the feedback of instability leads to a transverse broadening of the beam. Thus, $\alpha \leq 0.005$ represents the physical limit for observing these effects in laboratory conditions.
    \item In the laboratory-scale simulations, the diluted beam is losing around 29\% of its kinetic energy due to instability. Of this, approximately 7\% of the beam kinetic energy is contributing to the electromagnetic field growth. We find that the beam energy loss rate slows down after the linear growth phase ends, and the system continues to evolve non-linearly. Despite that, the non-linear feedback of the instability has only a marginal effect on the angular broadening of the beam.
    \item In a very dilute beam, electromagnetic instabilities are subdominant, allowing electrostatic oblique instabilities to take precedence. This leads to a decrease in the emerging magnetic field strength, suggesting that the rates of transverse beam broadening are also reduced. In astrophysical scenarios, the background intergalactic medium (IGM) has a density of approximately $n_{\text{IGM}} \sim 10^{-7} \text{cm}^{-3}$, which translates to the density contrast of $\alpha = 10^{-21} - 10^{-15}$ for lower redshift ($z<1$) TeV sources. This indicates that the beam is very dilute. The beam lost approximately $2-4\%$ of its energy due to instability. Moreover, the non-linear feedback from instability effects is almost marginal, leading to a negligible transverse broadening. 
    \item While preparing this paper, another study by \cite{Alawashra:2024ndp} was published that conducted a quantitative assessment of the blazar 1ES 0229+200 using an alternative numerical approach to investigate the angular spread of the beam induced by the instability. The study concluded that the angular broadening is very minimal. In a steady-state scenario, the angular spread of the beam is approximately $5\cdot10^{-6}~\text{rad}$ for a Lorentz factor of $10^{7}$, with a gamma-ray photon mean-free path of about $13~\text{kpc}$. However, for the same source, we note that our extrapolated estimation of the angular broadening of a beam with bulk Lorentz factor of $10^{7}$ in the non-linear phase (approaching saturation regime) is $\sim 6.7 \cdot 10^{-4}~\text{rad}$, which is about two orders of magnitude larger than what was estimated in \cite{Alawashra:2024ndp}.
\end{itemize}
In conclusion, we have identified the physical limit of the beam density contrast for which a warm beam can be inferred as "dilute," conducting real-life laboratory experiments that mimic the realistic astrophysical pair beam produced from TeV blazars.  Although we have studied the evolution of pair beams in an unmagnetized background plasma, it is also worthwhile to explore their behavior in a magnetized plasma, as this can suppress instability growth and modify the condition for dilute beams. Another important consideration can be collisional effects, particularly if the collisional frequency becomes comparable to the growth rate of the filamentation instability, which can further suppress the instability.
\section*{Data availability}
No data is used for the research described in the article.

\section*{CRediT authorship contribution statement}
\textbf{Suman Dey:} Writing – original draft, Writing – review \& editing, Software, Methodology, Investigation, Formal analysis, Data curation, Conceptualization, Visualization, Validation. \textbf{G\"{u}nter Sigl:} Writing – review \& editing, Methodology, Conceptualization, Supervision, Resources, Visualization, Validation.

\section*{Declaration of competing interest}
The authors confirm that they have no competing financial interests or personal relationships that could have influenced the work presented in this paper.

\section*{Acknowledgments}
SD was funded by the Deutsche Forschungsgemeinschaft (DFG, German Research Foundation) under Germany’s Excellence Strategy– EXC 2121 “Quantum Universe”– 390833306. This project was conceived by GS. The authors gratefully acknowledge the EPOCH code development team. The authors acknowledge the HPC facility of the Maxwell computational resources operated at Deutsches Elektronen-Synchrotron (DESY), Hamburg, Germany.  
The authors would like to thank Prof. Martin Pohl and Dr. Mahmoud Alawashra for the insightful discussions during the Zeuthen visit. We are grateful to the reviewer for the constructive input.

\appendix

\section{Maxwell-J\"{u}ttner distribution}\label{sec:composite-maxwell}
In this section, we compare the instability growth of a broad beam with a Cauchy distribution to that of a Maxwellian beam distribution.
\begin{equation}
    f(\textbf{p}; \mu, \sigma_{\parallel, \perp}) \propto e^{-\gamma\left[\left\{1 +\frac{\left(p_{x} - \mu\right)^{2}}{2\sigma_{\parallel,0}^{2}} + \frac{p_{y}^{2} + p_{z}^{2}}{2\sigma_{\perp,0}^{2}}\right\}^{1/2}\right]},
	\label{eqn:maxwell-fn}
\end{equation}
where $\gamma, \mu$ values are the same as used before for the Cauchy distribution. A comprehensive overview of the parameters for this type of beam distribution is provided in the same manner as for the Cauchy beam distribution, as shown in Table \ref{tab:ex-table1}.
\begin{figure}[ht]
\centering
\includegraphics[width=0.47\textwidth]{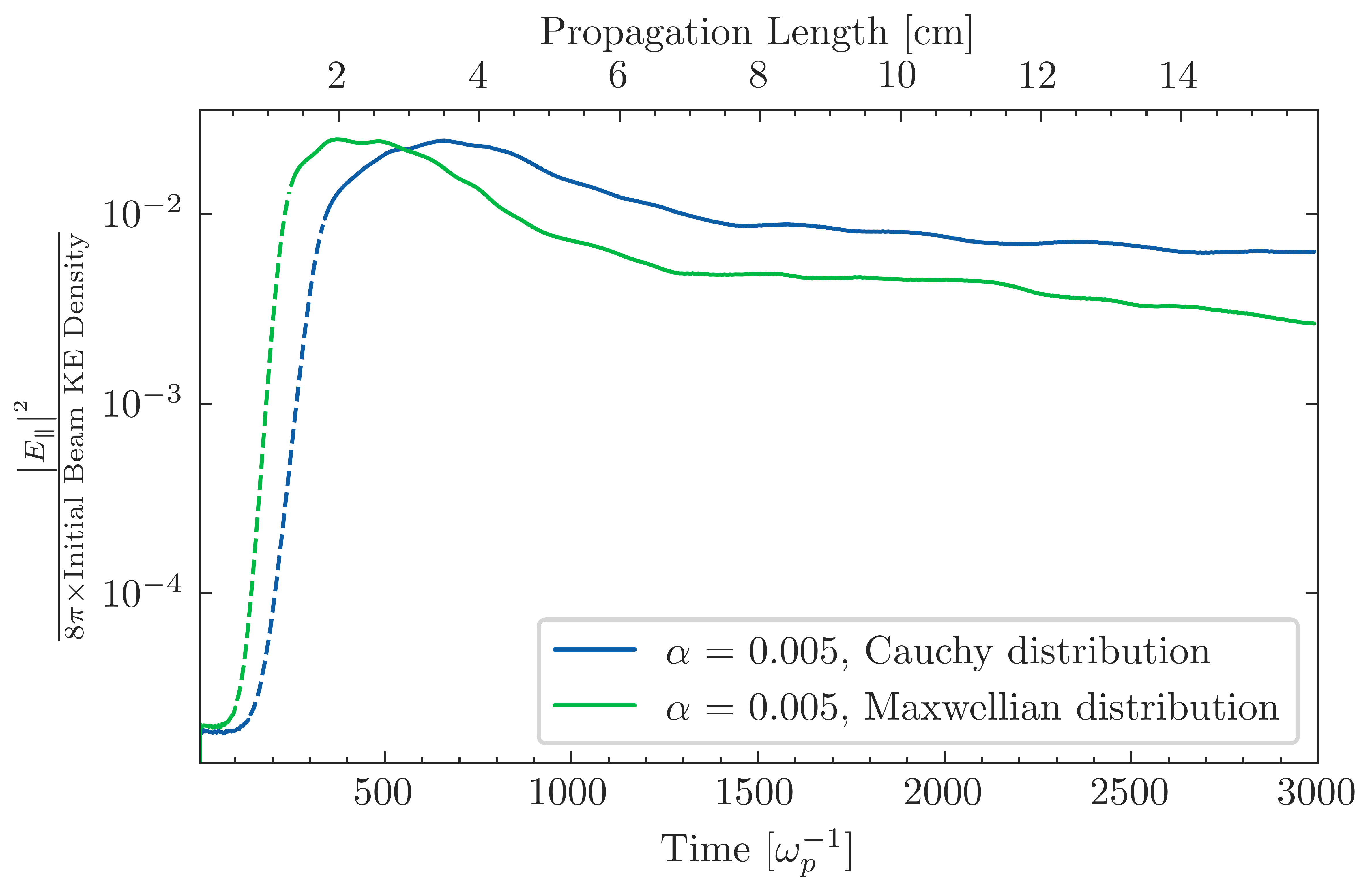}
\caption{
The time evolution of the fraction of initial beam kinetic energy is converted into the electric field for a broad Maxwellian and Cauchy beams, for $\alpha = 0.005$. The dashed lines represent the beginning and end of the instability growth phase.}
\label{fig:maxwell-compare}
\end{figure}
Fig.~\ref{fig:maxwell-compare} shows the fraction of beam kinetic energy converted into the electric field for Maxwellian beams from Eq.~(\ref{eqn:maxwell-fn}) in comparison to the Cauchy distribution from Eq.~(\ref{eqn:dist-fn}). The distance or time over which the instability develops is nearly the same; however, the impact of non-linear behavior for these two cases causes changes in the saturation level by factors of a few.

\section{Numerical convergence}\label{sec:numerical-convergence}
In any realistic simulation employing finite grid resolution and a finite number of macro-particles, it is essential to ensure the numerical convergence. In our simulations, plasma skin depth ($c\cdot \omega_{p}^{-1}$) is resolved with $8$ grid cells in both the longitudinal and transverse directions. However, the choice of the number of macro-particles used to sample the particle distribution function can lead to artificial numerical heating. For instance, using too few macro-particles per cell can result in large field fluctuations as particles move between cells. Therefore, it is essential to ensure that numerical heating remains under control in simulations. We have checked the stability of the growth rate of the instability for different macro-particles per cell. The growth rate remains stable when using more than $150$ macro-particles per cell. We verify that the total energy of the system is conserved to within about $<1\%$ accuracy, which improves with a sufficient number of macro-particles per cell. 
\biboptions{sort&compress}
\bibliographystyle{elsarticle-num} 
\bibliography{ref}

\end{document}